\documentclass[twocolumn]{aastex61}
\pdfoutput=1 %for arXiv submission
\usepackage{amsmath,amstext}
\usepackage[T1]{fontenc}
\usepackage{apjfonts} 
\usepackage[figure,figure*]{hypcap}
\usepackage{graphicx}
\usepackage{natbib}
\usepackage{epsfig}
\usepackage{epstopdf}
\usepackage{enumerate}
\usepackage{longtable} 
\usepackage{multirow, bigdelim}

\usepackage{color}
\usepackage[normalem]{ulem} 
\usepackage{xspace}
\definecolor{green2}{rgb}{0,0.5,0}
\definecolor{green2}{rgb}{0.2,0.6,0.1}
\definecolor{orange}{rgb}{1,0.5,0}

\newcommand{\push}{\mathrm{push}}
\newcommand{\msun}{M$_{\odot}$\xspace}
\newcommand{\kpush}{$k_{\rm push}$\xspace}
\newcommand{\trise}{$t_{\rm rise}$\xspace}

\shorttitle{Pushing 1D CCSNe to explosions: Nucleosynthesis yields}
\shortauthors{Curtis et al.}

\begin{document}

\title{PUSHing core-collapse supernovae to explosions in spherical symmetry III: Nucleosynthesis Yields}

\author{Sanjana Curtis}
\affiliation{Department of Physics, North Carolina State University, Raleigh NC 27695}
\author{Kevin Ebinger}
\affiliation{Department of Physics, North Carolina State University, Raleigh NC 27695}
\affiliation{GSI Helmholtzzentrum f\"ur Schwerionenforschung, D-64291 Darmstadt, Germany}
\author{Carla Fr\"ohlich}
\affiliation{Department of Physics, North Carolina State University, Raleigh NC 27695}
\author{Matthias Hempel}
\affiliation{Department f\"ur Physik, Universit\"at Basel, CH-4056 Basel, Switzerland}
\author{Albino Perego}
\affiliation{Istituto Nazionale di Fisica Nucleare, Sezione Milano Bicocca, Gruppo Collegato di Parma, I-43124 Parma, Italy}
\affiliation{Dipartimento di Fisica, Universi\`a degli Studi di Milano Bicocca, I-20126 Milano, Italy}
\affiliation{Dipartimento di Scienze Matematiche Fisiche ed Informatiche, Universit\`a di Parma, I-43124 Parma, Italy}
\author{Matthias Liebend\"orfer}
\affiliation{Department f\"ur Physik, Universit\"at Basel, CH-4056 Basel, Switzerland}
\author{Friedrich-Karl Thielemann}
\affiliation{Department f\"ur Physik, Universit\"at Basel, CH-4056 Basel, Switzerland}
\affiliation{GSI Helmholtzzentrum f\"ur Schwerionenforschung, D-64291 Darmstadt, Germany}
\email{ssanjan@ncsu.edu}
\email{cfrohli@ncsu.edu}

\begin{abstract}
In a previously presented proof-of-principle study, we established a parametrized spherically symmetric explosion method (PUSH) that can reproduce many features of core-collapse supernovae for a wide range of pre-explosion models. The method is based on the neutrino-driven mechanism and follows collapse, bounce and explosion. There are two crucial aspects of our model for nucleosynthesis predictions. First, the mass cut and explosion energy emerge simultaneously from the simulation (determining, for each stellar model, the amount of Fe-group ejecta). Second, the interactions between neutrinos and matter are included consistently (setting the electron fraction of the innermost ejecta).
In the present paper, we use the successful explosion models from \cite{push2} which include two sets of pre-explosion models at solar metallicity, with combined masses between 10.8 and 120~M$_{\odot}$. We perform systematic nucleosynthesis studies and predict detailed isotopic yields. The resulting $^{56}$Ni ejecta are in overall agreement with observationally derived values from normal core-collapse supernovae. The Fe-group yields are also in agreement with derived abundances for metal-poor star HD~84937. We also present a comparison of our results with observational trends in alpha element to iron ratios.
\end{abstract}

\keywords{nucleosynthesis, supernovae: general, supernovae: core-collapse}
\section{Introduction} \label{sec:introduction}

Core-collapse supernovae (CCSNe) are extremely energetic explosions that mark the deaths of massive stars, either via the collapse of an ONeMg-core (for stars with $M$ in the interval 8 - 10 $\rm{M}_{\odot}$) or via the collapse of an Fe-core (for $M >10$~M$_{\odot}$). CCSNe play a vital role in the synthesis and dissemination of many heavy elements in the universe. When massive stars die, they expel elements produced during the hydrostatic and explosive burning phases into the interstellar medium, thereby enriching host galaxies and driving chemical evolution.

Understanding the explosion mechanism of CCSNe is a longstanding problem which is still not fully solved. It is a complex problem that requires fluid dynamics, general relativity, a nuclear equation of state and neutrino physics, and in principle can only be solved in three dimensions. In addition, it also depends on the initial conditions, that is, on the pre-explosion properties of the massive star. For the most recent advances see, for example, \citet{nomoto06,heger10,chieffi13,nomoto13,chieffi17,hirschi17p,meynet17book,nomoto17}. The reviews by \cite{janka16} and \cite{Burrows2018} provide an overview of the status of CCSN modeling, while results related to nucleosynthesis in massive stars are discussed in \cite{nomoto13}.
A growing number of 2D and 3D CCSN simulations are currently available
\citep[e.g.][]{janka12,burrows13,nakamura15,janka16,bruenn16}, however the solution is not yet converged. Simulations in 2D generally lead to explosions more readily, but the resulting explosion energies may be higher in 3D \citep{takiwaki14,lentz15,melson15,janka16,hix16}.
However, in addition to the question of the explosion mechanism,
there exists a concurrent need to provide robust CCSN nucleosynthesis yields for progenitor stars of different masses and metallicities. These yields are required both for comparison with observations of metal-poor stars and for the fast progressing field of galactic chemical evolution (GCE).

To date, only very few nucleosynthesis predictions from multi-dimensional CCSN simulations are available. 
\cite{Harris.Hix.ea:2017} investigated the uncertainties in post-processing axisymmetric (2D) simulations of four pre-explosion models (12, 15, 20, and 25~M$_{\odot}$) and identified several sources of uncertainty impacting the final yields.
\cite{wanajo2d} recently carried out nucleosynthesis calculations for 2D simulations of neutrino-driven supernova explosions. However, the $^{56}$Ni masses obtained in their study are only lower limits for the massive models, for which they omitted the outer envelopes and large parts of the Si/O layer. 
\cite{Eichler.Nakamura.ea:2017} focused on the production of light p-nuclei in two 2D CCSN simulations (11.2 and 17.0~M$_{\odot}$). \cite{yoshida2017} investigated nucleosynthesis in 2D neutrino-driven explosions of ultra-stripped type Ic supernovae (1.45 and 1.5~M$_{\odot}$ CO stars), finding that elements between Ga and Zr are produced in the neutrino-irradiated ejecta, although the total yields were not clearly determined. Electron-capture supernovae were studied in 2D \citep{2011wanajo, 2013awanajo,2013bwanajo}, using an ONeMg core that emerged from the evolution of an 8.8~M$_{\odot}$ SAGB star. The simulations were post-processed to examine the production of various neutron-rich isotopes, light trans-iron elements, and weak r-process elements.
A 3D simulation of the neutrino-driven explosion of a 15~M$_{\odot}$ red supergiant was carried out by \cite{2017wong} using a parametrized neutrino engine. They compared the resulting spatial distribution of radioactive nuclei in their model to the morphology of the Cas~A remnant, concluding that a neutrino-driven explosion can account for the observed asymmetries.

The explosion mechanism problem, combined with the high computational costs of multi-dimensional simulations, warrants the use of alternative methods for answering nucleosynthesis related questions.
A general picture of CCSN nucleosynthesis is well-established.
The outer layers ejected in supernova explosions largely contain the products the hydrostatic burning phases from stellar evolution, which are only mildly destroyed during the explosion. They contain the elements He, C, O, Ne, and Mg (from H, He, and C-burning). Intermediate layers are characterized by some explosive C and Ne-burning (adding/replacing some O, Mg, and also Si to/of the initial hydrostatic yields, which were destroyed during their explosive processing). Layers further in experience explosive O-burning (producing Si, S, and also Ca and Ar), while the innermost layers experience explosive Si-burning that has $^{56}$Ni as the major burning product. However, it can be realized that some He is retained in the inner ejecta. This is due to explosive Si-burning with a so-called alpha-rich freeze-out. In the shock, most matter is first disintegrated into neutrons, protons, and alpha-particles, and the build-up of heavier nuclei occurs only during the subsequent cooling and expansion. If the densities are too low, the triple-alpha reaction and competing reactions (for example $\alpha\alpha n$), which eventually pass the bottle neck to carbon production, are not effective enough to permit all matter to reach nuclei in the Fe-group with the highest binding energy, such as $^{56}$Ni. Such an alpha-rich freeze-out of matter that initially attained nuclear statistical equilibrium will also retain alpha-nuclei along the chain to $^{56}$Ni, such as $^{32}$S, $^{36}$Ar, $^{40}$Ca, $^{44}$Ti. $^{48}$Cr, and $^{52}$Fe. These nuclei can be found in the inner ejecta, where their final abundances correlate with the final He abundance. The final He abundance in turn is related to $T^3/\rho$, a quantity proportional to the entropy of a radiation-dominated plasma. Thus, a larger entropy of matter during explosive Si-burning stands for a stronger alpha-rich freeze-out, and thus higher abundances of these nuclei, in tandem with higher He-abundances.

For the intermediate stellar layers, the nucleosynthesis mainly depends on the strength of the shock wave. There, methods that deposit a prescribed amount of energy into the pre-explosion model work quite well.
Early attempts at such artificially triggered explosions for CCSN nucleosynthesis are pistons \citep{woosley95,rauscher02,hw07,heger10}, thermal energy bombs \citep{thielemann96,nomoto06,umeda08,nomoto13,nomoto17}, and kinetic energy bombs \citep{limongi06,limongi12,chieffi13,chieffi17}.
The piston method specifies the ballistic trajectory of a radial mass zone such that a typical final explosion energy of $E=1.2 \times 10^{51}$~erg is achieved. In this method, the mass cut between the neutron star and the ejecta is specified by the chosen position of the piston and an entropy value.
In the thermal and kinetic energy bomb methods, a prescribed amount of energy is deposited in a deep layer of the pre-explosion model. The mass cut is obtained by integrating the nucleosynthesis yields from the outside inwards, until a typical amount of the order of 0.1~M$_{\odot}$ of $^{56}$Ni is reached.
In these traditional approaches, the physics of the explosion phase is not included and the explosion energy is not determined self-consistently with the mass cut between the proto-neutron star (PNS) and the ejecta. Since the mass cut and explosion energy are effectively treated as free parameters, the predictive power of these methods is limited. 
When \cite{romano2010} quantified the uncertainties introduced by the choice of stellar yields in galactic chemical evolution models, they noted that varying the mass cut location in models of CCSN explosions has a huge impact on GCE modeling. These uncertainties are especially pronounced for the majority of the iron-peak elements.

The traditional methods are therefore not suitable for the innermost layers, where the nucleosynthesis is directly related to the details of the explosion and the electron fraction, $Y_e = \langle Z/A \rangle$, of the material.
Important aspects which affect the composition of ejected matter (for all type of simulations, multi-D and spherically symmetric) are related to weak interactions. 
High electron densities, caused by high Fermi energies of (degenerate) electrons, lead to neutron-rich conditions due to electron captures on free protons during the collapse with a proton/nucleon ratio $Y_e<0.5$. The $Y_e$ of the material is strongly affected during the explosion by neutrino and anti-neutrino captures on protons and neutrons, governed by the reactions: 
$\nu_e + n  \leftrightarrow p+ e^-$ and $\bar\nu_e + p \leftrightarrow n+ e^+$.
These reactions turn matter neutron-rich if the average anti-neutrino energy $\langle\epsilon_{\bar\nu_e}\rangle$ is higher than the average neutrino energy $\langle\epsilon_{\nu_e}\rangle$ by 4 times the neutron-proton mass difference $\Delta$ for similar (electron) neutrino and anti-neutrino luminosities \citep{qian96}. Thus, in most cases the first reaction (being energetically favorable) wins, changing $Y_e$ from the initial neutron-rich conditions towards values beyond $Y_e=0.5$.
The traditional methods such as the piston or thermal/kinetic bomb methods ignored such interactions between matter and neutrinos. \citet{cf06a} have shown that including the neutrino interactions affects the iron group yields, in particular of $^{45}$Sc and $^{64}$Zn, and also leads to the $\nu p$-process \citep{cf06b,  pruet06, 2011wanajo}.

The next logical step are methods that try to mimic the multi-dimensional neutrino heating in spherical symmetry.
In the ``neutrino light-bulb method'', this is achieved by excising the PNS and replacing it with an inner boundary condition, which uses an analytical prescription for the neutrino luminosities. Successful explosions can be achieved for suitable choices of neutrino luminosities and energies \citep[see e.g.][]{yamasaki05,iwakami08,yamamoto13}.
In ``absorption methods'' \citep{cf06a,cf06b,fischer10}, this is accomplished by modifying the neutrino-opacities in models with full Boltzmann neutrino transport to obtain increased energy deposition and successful explosions.

Recently, various groups have attempted to mimic the net effects of multi-dimensional neutrino transport in ways that are adapted more consistently to core collapse and PNS accretion.
\cite{Ugliano.Janka.ea:2012} used a sophisticated light-bulb method where they obtain explosions in spherical symmetry by neutrino energy deposition in post-shock layers. They replace the innermost 1.1~M$_{\odot}$ of the PNS by an inner boundary and use gray neutrino transport. An analytic cooling model is used for the evolution of the neutrino luminosity at the inner boundary. 
The Prometheus-Hot Bubble (P-HOTB) approach was used to predict explosion energies and neutron star masses as function of stellar mass \citep{ertl16} and nucleosynthesis yields \citep{sukhbold16}. To predict nucleosynthesis, the free parameters were set by fitting to observational properties of SN~1987A for progenitor masses around 20~M$_{\odot}$ and to properties inferred from the Crab SN remnant for a 9.6~M$_{\odot}$ progenitor. However, \citet{sukhbold16} did not include the effects of neutrino interactions on $Y_e$ in the nucleosynthesis.

The PUSH method is a different approach that also includes the weak interactions (especially the neutrino-nucleon interactions) that affect the electron fraction $Y_e$.
\cite{push1} --- from here on Paper~I --- utilized the muon and tau neutrinos as an additional energy source that approximately captures the essential effects  that influence the (net) neutrino heating seen in multi-dimensional simulations.  
A fraction of the luminosity of heavy-flavor ($\nu_x$) neutrinos emitted by the PNS is deposited behind the shock in order to re-energize it. This is inspired by the increased neutrino-heating that fluid elements experience in multi-dimensional simulations.  By using the $\nu_x$ properties, dynamical feedback from the accretion history and cooling properties of the PNS are included self-consistently. It also allows us to trigger
explosions without modifying the electron-flavor neutrino luminosities nor changing the charged current reactions. This increases the accuracy of the electron fraction in the innermost ejecta, which is crucial for nucleosynthesis predictions. In addition, the mass cut emerges naturally in the PUSH method. Both aspects are critical for accurate nucleosynthesis predictions, in particular of the Fe-group elements. In Paper~I, we showed that the PUSH method is capable of causing explosions in spherically-symmetric CCSN simulations and that the method can reproduce the well-known properties of SN~1987A (explosion energy and yields of $^{56-58}$Ni and $^{44}$Ti) for a suitable pre-explosion model ($18 \leqslant M_{\mathrm{ZAMS}} \leqslant 21$~M$_{\odot}$) and parameter choices.

In \cite{push2} --- from here on Paper~II --- we identified three observational constraints on CCSNe that we use to set the free parameters in the PUSH method:
(i) reproduce the observed properties of SN~1987A for a suitable pre-explosion model,
(ii) allow for the possibility of black-hole formation,
and
(iii) result in lower explosion energies for stars with ZAMS masses below $\lesssim 13$~M$_{\odot}$ (``Crab-like SNe'').
We applied this improved PUSH method to two samples of solar metallicity pre-explosion models from \cite{Woosley.Heger:2002} and \cite{hw07}. In Paper~II, we discuss the explosion properties and the progenitor-remnant connection.

Here, we want to exploit the predictive nature of the PUSH method for nucleosynthesis predictions. It is important to note that with our approach, only one model (the ``SN~1987A progenitor'' -- model s18.8 in this study) does not have predictive power. For all other models, we can predict complete nucleosynthesis yields, including Fe-group elements, from our approach.

The organization of this article is as follows. We describe the input and explosion models as well as the nuclear reaction network in Section \ref{sec:input}. Section \ref{sec:results} contains a discussion of the nucleosynthesis trends and comments on the synthesis of various isotopes. In Section \ref{sec:observations}, we present comparisons of our results with observations of CCSNe and metal-poor stars. A summary is given in Section \ref{sec:summary}.

\section{Input models and computational procedure} \label{sec:input}
\subsection{Initial models} \label{subsec:progenitors}

We explore the nucleosynthesis yields of two sets of pre-explosion models: WHW02 and WH07, presented in \cite{Woosley.Heger:2002} and \cite{hw07} respectively. These are spherically symmetric, non-rotating, solar metallicity models from the stellar evolution code {\tt KEPLER}.
We use the same naming convention as in Paper~II, where the pre-explosion models are labeled by their zero-age main sequence (ZAMS) masses and a letter indicating to which series they belong (see Table~\ref{tab:progenitors} in this Paper). The ZAMS masses lie in the 10.8 -- 75.0~M$_{\odot}$ range for models in the WHW02 set and in the 12.0 -- 120.0~M$_{\odot}$ range for those in the WH07 set.
We only predict nucleosynthesis yields for the models that resulted in successful explosions when the PUSH method was applied; failed explosions leading to the formation of black holes are excluded from the present study as they are not expected to eject a significant amount of explosive nucleosynthesis products \citep{lovegrove13,lovegrove17}.

\begin{table}  
\begin{center}
	\caption{Pre-explosion models.
    	\label{tab:progenitors}
	}
	\begin{tabular}{lllllc}
	\tableline \tableline 
Series & Label & Min Mass & Max Mass & $\Delta m$ & Ref. \\
 &  & (M$_{\odot}$) & (M$_{\odot}$) & (M$_{\odot}$) &  \\
	\tableline
WHW02 & s & $10.8$ & $22.0$ & $0.2$ & 1 \\
	  &   & $22.4$ & $22.6$ & $0.2$ & 1 \\
	  &   & $25.8$ & $28.2$ & $0.2$ & 1 \\
	  &   & $29.0$ & $38.0$ & $1.0$ & 1 \\
	  &   & $40.0$ &        &       & 1 \\
	  &   & $75.0$ &        &       & 1 \\
WH07 & w & $12.0$ & $22.0$ & $1.0$ & 2 \\
	 &   & $25.0$ & $33.0$ & $1.0$ & 2 \\
	 &   & $50.0$ & $60.0$ & $5.0$ & 2 \\
     &   & $70.0$ & & & 2 \\
     &   & $80.0$ & $120.0$ & $20.0$ & 2 \\
	\tableline
	\end{tabular}
\end{center}
\tablecomments{All models have solar metallicity. Note that this Table only contains models that explode, not the entire pre-explosion sets. This corresponds to 84 models from the WHW02 set and 27 models from the WH07 set.}

\tablerefs{(1)~\citet{Woosley.Heger:2002}; (2)~\citet{hw07}}
\end{table}

\begin{figure}
\includegraphics[width=0.49\textwidth]{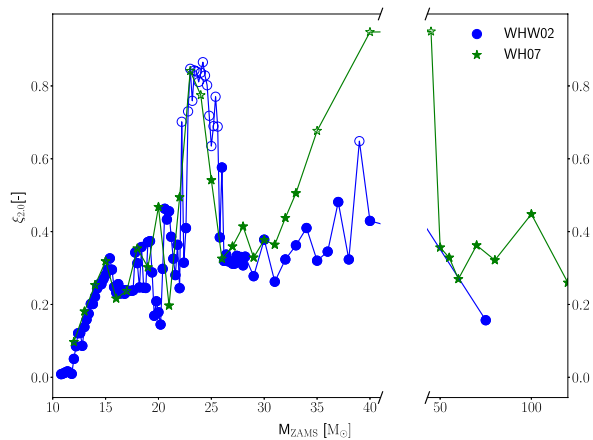}
	\caption{Compactness  $\xi_{2.0}$ (calculated at 2.0~M$_{\odot}$) as function of the ZAMS mass of pre-explosion models. Both sets of initial models are shown: WHW02 (blue circles) and WH07 (green stars). Open symbols represent models that formed black holes instead of exploding and are thus excluded from our study.
		\label{fig:series_compact}
     }
\end{figure}

It is well-known that the ZAMS mass is not a good predictor of the final fate of a stellar model. Here, we make use of the compactness parameter as defined by \citet{OConnor.Ott:2011} to describe the pre-explosion models by a single number. The compactness is defined as the ratio of a given mass $M$ and the radius $R(M)$ which encloses that mass:
\begin{equation}
\xi_{M} \equiv \frac{M/M_{\odot}}{R(M)/1000\mathrm{km}},
\label{eq:compactness}
\end{equation}
where we use $M=2.0$~M$_{\odot}$. With this choice of mass, the compactness characterizes the innermost structure of the pre-explosion model, which is expected to strongly influence the explosion dynamics and hence also the nucleosynthesis. 
Consistent with Paper~II, we compute the compactness at bounce. Figure~\ref{fig:series_compact} shows the compactness of all models (WHW02 in blue, WH07 in green).  The relationship between compactness and ZAMS mass is not monotonic due to mass loss and due to the details of the evolution of burning shells before collapse, as is evident in Figure \ref{fig:series_compact} and as has been extensively discussed by \cite{sukhbold.woosley:2014} and \cite{sukhbold17}. Note that the lowest-mass models included in this study have very low compactness values. This is because 2~M$_{\odot}$ (where we evaluate the compactness) encloses not only the core but also parts of the He-layer for these models.

\subsection{Explosion models}
\label{subsec:explosions}

For the explosion simulations, we use the hydrodynamics code Agile together with spectral neutrino transport. We employ IDSA for electron-flavor neutrinos \citep{Liebendoerfer.IDSA:2009} and ASL for heavy-flavor neutrinos \citep{perego16}. We use the HS(DD2) nuclear equation of state for matter in NSE conditions \citep{Hempel.SchaffnerBielich:2010,typel10} and an extension to non-NSE conditions using an ideal gas approach coupled to an approximative alpha-network \citep{push1}. 

The PUSH method triggers explosions by artificially enhancing the neutrino heating with a local heating term
\begin{equation}   
   Q^+_{\push} (t,r) = 4 \, \mathcal{G}(t) \int_0^{\infty} q^+_{\push}(r,E) \, dE ,
   \label{eq:push_integral}
\end{equation}
where
\begin{equation}   
   q^+_{\push}(r,E) \equiv 
   \sigma_0 \;
   \frac{1}{4 \, m_b} 
   \left( \frac{E}{m_e c^2} \right)^2 
   \frac{1}{4 \pi r^2} 
   \left( \frac{dL_{\nu_x}}{dE} \right)
   \mathcal{F}(r,E). 
   \label{eq:push_qdot}
\end{equation}
Here, $E$ is the energy and $(dL_{\nu_x}/dE)/(4 \pi r^2)$  is the spectral energy flux of any single neutrino species $\nu_x$, $\sigma_0$ is a typical neutrino cross-section, and $m_b$ is the average baryon mass. All the four parameters of the PUSH method are contained in the temporal evolution function $\mathcal{G}(t)$. Two of the parameters ($t_{\mathrm{on}}$ and $t_{\mathrm{off}}$) are robustly set (see Paper~I). The other two parameters ($k_{\mathrm{push}}$ and $t_{\mathrm{rise}}$) are free parameters that are calibrated by comparing the obtained explosion properties with observational data of supernovae and to results of multi-dimensional simulations (see Paper~II for details on how this was done). 

Paper~II discusses two possible calibrations: the standard calibration uses compactness $\xi_{2.0}$ while the second calibration uses $\xi_{1.75}$.
On the one hand, with the standard calibration, a broader set of models explode successfully, allowing us to investigate the nucleosynthesis outcomes over a potentially wider range of progenitor masses.
On the other hand, the second calibration depends more strongly on the mass of the iron core and results in a larger fraction of faint explosions and black holes, especially for high mass progenitors (M > 20.0 M$_{\odot}$). Thus, for our analysis we use the parameters from the standard calibration of Paper~II.

In the standard calibration, the value of \trise is fixed at $t_{\mathrm{rise}}=400$~ms and \kpush has compactness-dependent values given by $k_{\mathrm{push}}(\xi)=a\xi^2+b\xi+c$ where $a=-23.99$, $b=13.22$, and $c=2.5$. This sets the free parameters for all models \emph{a priori} and gives the method a predictive nature. 
We refer the reader to Paper~I and Paper~II for the details on the method, the calibration procedure, explosion properties, and the progenitor-remnant connection.

The PUSH setup enables a consistent treatment of three aspects critical for reliable nucleosynthesis predictions: the location of the mass cut, the electron fraction of the ejecta, and the entropy.
We include all the matter from the center to the He-layer in our computational domain (excluding only the low-density H-envelope), with the exception of the low mass model s11.2 and a few models with ZAMS masses above 30~M$_{\odot}$ with large pre-explosion mass loss.
This allows us to follow the formation of the PNS during collapse and its subsequent evolution as the explosion develops and succeeds. The (in 1D, simple) bifurcation between ejected matter and matter settling on the PNS emerges naturally in our simulations and is consistent with the explosion energy. For nucleosynthesis predictions, the location of the mass cut is crucial in setting the amount of Fe-group ejecta which originate from the innermost ejected layers. 

The detailed composition of the innermost ejecta depends critically on the electron fraction $Y_e$ of these layers \citep{thielemann96,cf06a}. In these regions, the weak processes during collapse, bounce, and explosion significantly alter the electron fraction compared to its pre-collapse value. In particular, the electron-flavor neutrinos and anti-neutrinos play a crucial role in setting the relative ratio of neutrons to protons through the charged-current reactions on free nucleons \citep{cf06a,pruet06,wanajo06}:
\begin{eqnarray}
n + \nu_e &\leftrightarrow& e^- + p \\
p + \bar{\nu}_e &\leftrightarrow& e^+ + n.
\end{eqnarray}
We note that, similar to multi-D simulations, neutrino flavor oscillations are not included, which in principle may alter the electron fraction \citep{Mirizzi.ea:2016,Horiuchi.Kneller:2018}.
We want to stress that unlike earlier attempts at using neutrinos in 1D simulations to obtain explosions \citep{cf06a,fischer10}, the PUSH approach does not modify the transport of electron-flavor neutrinos. IDSA contains the most relevant neutrino-nucleon reactions that set the $Y_{e}$ in the innermost ejecta. These include electron neutrino and anti-neutrino captures on nucleons, neutrino scattering on nucleons and nuclei, and pair production.

All pre-explosion models considered here were exploded with the PUSH method and the parameter setting described above. More detail on the explosion dynamics and outcomes can be found in Paper~II.

%%%%%%%%%%%%%%%%%%%%%%%%%%%%%%%%%%%%%%%%%%%%%%%%%%%
\subsection{Nucleosynthesis calculations}
\label{subsec:network}

The detailed yields for all exploding models are calculated using a post-processing approach. 
For the nucleosynthesis post-processing, we use the nuclear reaction network {\sc CFNET}. The network includes up to 2902 isotopes (depending on the pre-explosion model), covering free nucleons as well as neutron-deficient and neutron-rich isotopes up to $^{211}$At.
We use the reaction rate library REACLIB \citep{reaclib}:
the reaction rates included are based on experimentally known rates wherever available. Where experimental rates do not exist, the n-, p-, and alpha-captures are taken from \citet{Rauscher.FKT:2000}. 
We also include weak interactions (which have an impact on the electron fraction), i.e., 
electron and positron capture rates from \cite{lmp},
$\beta^-$/$\beta^+$ decays from the nuclear database \textit{NuDat2}\footnote{http://www.nndc.bnl.gov/nudat2/} where available and from \cite{Moller.ea:1995} otherwise, and
neutrino and anti-neutrino captures on free nucleons.

For the post-processing, the ejecta of each model are divided into radial mass elements of $10^{-3}$~\msun each (referred to as `tracer particles' or `tracers'). 
For each tracer particle, we obtain its hydrodynamic and local neutrino history from the start to the last time slice of the explosion simulation (approximately $\sim 4.6$~s post-bounce).
We post-process all tracers that reach a peak temperature $\geqslant 1.75$~GK. If the tracer particles reach a peak temperature $\geqslant 10$~GK, we start the nucleosynthesis calculations at 10~GK, when the temperature starts dropping below that value. 
For these tracer particles, the initial abundances are assumed to be NSE abundances set by the density and $Y_e$ at 10~GK (taken from the hydrodynamic simulation). 
For all other tracers, we start our calculations at the beginning of the hydrodynamic simulation. The initial abundances are obtained from the approximate alpha-network implemented within it (WHW02), or from the pre-explosion model (WH07). The initial electron fraction at the beginning of the post-processing is the initial value from the hydrodynamical simulation.
In both cases, the electron fraction is evolved in the network consistent with the nuclear reactions occurring.
The nucleosynthesis calculation for all tracers ends when the tracer temperature drops below 0.05~GK.

For many tracers, in particular the innermost zones, the temperature and density at the end of the hydrodynamic simulation (and hence at the end of the tracer particle's history) are still high enough for nuclear reactions to occur. Therefore, we employ the standard approach of extrapolating the hydrodynamic history of the tracer particle as follows:
\begin{eqnarray}
r(t) &=& r_{\rm final}  + t v_{\rm final} \label{eq:extrapol_rad}, \\
\rho(t) &=& \rho_{\rm final} \left( \frac{t}{t_{\rm final}} \right)^{-3}, \\
T(t) &=& T[s_{\rm final},\rho(t),Y_e(t)] \label{eq:extrapol_t9},
\end{eqnarray}
where $r$ is the radial position, $v$ the radial velocity, $\rho$ the density, $T$ the temperature, $s$ the entropy per baryon, and $Y_e$ the electron fraction of the mass zone.  The subscript ``final'' indicates the final end time of the hydrodynamical simulation. Equations~(\ref{eq:extrapol_rad})~--~(\ref{eq:extrapol_t9}) correspond to a free expansion for the density and an adiabatic expansion for the temperature (see, for example, \cite{korobkin2012}). The temperature is calculated at each time-step using the equation of state of \citet{Timmes.Swesty:2000}.

\section{Nucleosynthesis results} \label{sec:results}

We have computed detailed nucleosynthesis yields for all 84 exploding models in the WHW02 series and all 27 exploding models in the WH07 series. For three select models --- s16.0, s18.8 (best fit to SN~1987A) and s21.0 --- the isotopic yields are given in Appendix~\ref{appx:tables}, Table~\ref{tab:finabs02}. 
Detailed isotopic yields for \emph{all} exploding models of the WH07 set are available electronically. 
A sample is shown in Appendix~\ref{appx:tables}, Table~\ref{tab:finabw07}. 
Additionally, yields of long-lived radionuclides are available as a separate table. A sample is shown in Appendix~\ref{appx:radionuclides} in Table~\ref{tab:radiosample}.

In the following, we first discuss the synthesis pathways and then we focus on two representative models (s16.0 and s21.0). Finally, we highlight some important trends of the nucleosynthesis yields with other model properties.

\subsection{Isotopic Yields from Massive Stars} \label{subsec:yields}

\subsubsection{Late neutrino wind ejecta}
\label{subsec:innermost-tracers}

We have used a mass resolution of $10^{-3}$~M$_{\odot}$ for the nucleosynthesis post-processing of our models. This is a rather coarse resolution for late-time neutrino-driven wind ejecta and hence the details are quite uncertain (see also Paper~I). In our models, the wind includes of the order of a hundredth of a solar mass of neutron-rich material immediately above the mass cut. In both representative models (s16.0 and s21.0), the electron fraction of this wind is around $\sim$0.42, with values as low as 0.35 for material closest to the mass cut, and with velocities of $\sim 10^9$~cm~s$^{-1}$. These conditions are not sufficient for a full r-process \citep[e.g.][]{farouqi2010}. However, we see some production of elements up to mass number $A \approx 140$ in these very late, neutron-rich wind ejecta. 
One should note that neutrino-electron scattering is not included in our neutrino transport, which is an important source of thermalization and down-scattering,
especially for the high-energy electron anti-neutrinos at late times, see \citet{fischer12}. This could significantly alter the electron fraction in the neutrino wind.
A more detailed analysis is required, which is beyond the scope of this paper.

\subsubsection{Iron group elements - Sc through Zn - and the $\nu$p-process}
\label{subsec:Fegroup_elements}

Core-collapse supernovae make an important contribution to iron-group species, even though the dominant production of these elements occurs in Type Ia (thermonuclear) supernovae. The iron-group elements are made in the inner layers of the star when these layers undergo complete or incomplete silicon burning. The yields of these elements are thus quite sensitive to the explosion details. The location of the mass cut, for example, determines how much of the silicon-oxygen shell that undergoes explosive burning is ultimately ejected. 
We find a marked difference between the electron fraction in the innermost layers at the end of our simulations and the original electron fraction of the pre-explosion model. The details of these effects vary with the pre-explosion models and also depend on the details of the explosion mechanism. However, some qualitative comments can be made. We consistently find a proton rich ($Y_e > 0.5$) environment close to the mass cut (above the late neutron-rich wind), typically including $\sim 8 \times 10^{-3}$~\msun of material. 
This affects the final abundances of some iron group elements, and it allows for some synthesis of proton-rich isotopes beyond iron through the $\nu$p-process. However, the total ejected yields of trans-iron material are dominated by contributions from the neutron-rich wind (discussed in Section~\ref{subsec:innermost-tracers}) and also have a contribution from the pre-explosion composition.
In a small region further out, the $Y_e$ is slightly lower than its pre-explosion value. Outside of this, the $Y_e$ remains unaffected by the explosion. 
The electron fraction in the innermost layers has an impact on the relative production of many members of the iron group, quite pronounced in certain cases.
Both $^{56}$Ni and $^{57}$Ni, for example, are produced in the neutron-rich layers where alpha-rich freeze-out occurs. Lower $Y_e$ values favor the formation of $^{57}$Ni over $^{56}$Ni, changing the $^{57}$Ni/$^{56}$Ni ratio (observed via $\gamma$-rays). 

Table \ref{tab:fegp_synth} gives the nucleosynthesis processes responsible for explosive synthesis of the iron-group elements. For each element, we list all the stable isotopes and also identify, in bold, the isotope(s) that typically make(s) the dominant contribution to the total yield of each element.

\subsubsection{Intermediate mass elements - O through Ca}
\label{subsec:intermed_elements}

The final yields of intermediate mass elements include significant contributions from hydrostatic as well as explosive burning processes. Details related to their synthesis are given in Table~\ref{tab:intermed_synth}. 
During the explosion, the intermediate mass elements are produced primarily through explosive oxygen burning, which consumes only a fraction of the $^{16}$O present in the progenitor star. However, a large portion of the oxygen-neon shell usually does not get hot enough for explosive processing (T $\sim$  3-4~GK) and is ejected essentially unaltered. Thus, the yields of elements such as O, Ne, Mg include contributions both from explosive as well as hydrostatic burning (dominated by the hydrostatic contribution).
Observationally, the general tendency found for average CCSN ejecta (i.e. integrated over a progenitor  mass distribution) is increased [$\alpha$/Fe] ratios in the range 0.3--0.5 ($\alpha$ standing for the so-called alpha-elements O, Ne, Mg, Si, S, Ar, Ca, Ti) and on average 0.1~M$_\odot$ of $^{56}$Ni (decaying to Fe). 
Although we do not expect the same total yields from core-collapse supernovae with different progenitor masses, the ratio of alpha elements to iron can be quite similar, as seen in the small scatter of [$\alpha$/Fe] in galactic evolution, even at the lowest metallicities. These trends will be explored in detail in a future study.

\subsubsection{Light elements - H through N}
For the production of the light elements (e.g.\ H, He, C), the details of stellar evolution (e.g.\ convective boundary mixing or rotation), hydrostatic burning, and mass loss set the yields (Li, Be and B have significant contributions from other sources, such as classical novae, low mass stars and/or cosmic ray spallation).
In addition to affecting the $Y_e$ of the innermost ejecta, neutrinos can also alter the abundances of some rare isotopes in the outermost layers of the star through inelastic neutral-current neutrino scattering {(``$\nu$-process'', \citet{woosley.hartmann.ea:1990}; \citet{heger2005}). Here, we did not include neutral-current neutrino reactions in our post-processing network.
Since the light element abundances remain essentially unchanged by the explosion, pre-explosion models that provide the detailed abundances of these elements are necessary for accurate predictions of total ejected yields.

\begin{table*}
	\begin{center}
		\caption{Explosive synthesis of iron-group elements (Sc-Zn)}
        	\label{tab:fegp_synth}
		\begin{tabular}{lllll}
			\tableline \tableline
			Element & Stable Isotopes & Made As &  Production \\
			\tableline
			Sc &  $^{45}$Sc & $^{45}$K, $^{45}$Ca, $^{45}$Sc, \textbf{$^{45}$Ti}  & Mostly as $^{45}$Ti in inner layers with $Y_e$ > 0.5. \\
			Ti & $^{46}$Ti & $^{46}$Ti & Mostly as itself in $Y_e$ > 0.5 zone.\\
               & $^{47}$Ti & $^{47}$Ca, $^{47}$Sc, $^{47}$Ti, \textbf{$^{47}$V }  &  Mostly as $^{47}$V in $Y_e$ > 0.5 layers during complete Si burning. \\
               & \textbf{$^{48}$Ti} & $^{48}$Ca, $^{48}$Ti, \textbf{$^{48}$Cr}  & Primarily as $^{48}$Cr in Si burning zones.   \\
               & $^{49}$Ti & $^{49}$Ca, $^{49}$Sc, $^{49}$V, \textbf{$^{49}$Cr} & Mostly as $^{49}$Cr in $Y_e$ > 0.5 zone during complete Si burning. \\
               &  $^{50}$Ti & $^{50}$Sc, \textbf{$^{50}$Ti} & Very small amounts in n-rich layers close to the mass cut.\\
            V  &  $^{51}$V &  $^{51}$Cr, \textbf{$^{51}$Mn}  &  Mostly as $^{51}$Mn in $Y_e$ > 0.5 layers during complete Si burning. \\
			Cr & $^{50}$Cr &  $^{50}$Cr     &  In $Y_e$ > 0.5 layers and Si burning zones.\\
			   & \textbf{$^{52}$Cr} & $^{52}$Cr, \textbf{$^{52}$Fe}   &       Primarily as $^{52}$Fe in Si burning zones. \\
			   & $^{53}$Cr & $^{53}$Mn, \textbf{$^{53}$Fe}  & Both are produced in  $Y_e$ > 0.5 layers and incomplete Si burning zone.\\
			   & $^{54}$Cr &  \textbf{$^{54}$Cr}, $^{54}$Mn &   Very small amounts in n-rich layers close to the mass cut.\\
            Mn &  $^{55}$Mn & $^{55}$Fe, \textbf{$^{55}$Co} & Primarily as $^{55}$Co in $Y_e$ > 0.5 layers and incomplete Si burning zone.        \\
			Fe & $^{54}$Fe &  $^{54}$Fe    & In $Y_e$ > 0.5 layers and incomplete Si burning zone.\\
               & \textbf{$^{56}$Fe} & $^{56}$Mn, $^{56}$Fe, $^{56}$Co, \textbf{$^{56}$Ni}  &  Primarily as $^{56}$Ni in Si burning zones. \\
			   & $^{57}$Fe &  $^{57}$Fe, $^{57}$Co, \textbf{$^{57}$Ni} & Primarily as $^{57}$Ni in complete Si burning zone. \\
			   & $^{58}$Fe & $^{58}$Fe & Very small amounts in n-rich layers close to the mass cut. \\
            Co &  $^{59}$Co & $^{59}$Fe, \textbf{$^{59}$Ni}, \textbf{$^{59}$Cu} & Both $^{59}$Ni and $^{59}$Cu are made in the complete Si burning zone.\\
			Ni & \textbf{$^{58}$Ni} &  $^{58}$Ni    & As itself in Si burning zones. \\
			   & \textbf{$^{60}$Ni} & $^{60}$Fe, $^{60}$Ni, \textbf{$^{60}$Cu}, \textbf{$^{60}$Zn} &  Mostly as $^{60}$Zn and $^{60}$Cu in complete Si burning.\\
			   & $^{61}$Ni & $^{61}$Ni, $^{61}$Co, \textbf{$^{61}$Cu}, \textbf{$^{61}$Zn} & Mostly made as $^{61}$Cu and $^{61}$Zn in the complete Si burning zone. \\
			   & $^{62}$Ni & $^{62}$Fe, $^{62}$Co, \textbf{$^{62}$Ni}, \textbf{$^{62}$Zn} & As $^{62}$Zn during complete Si burning, as $^{62}$Ni close to mass cut.\\
			   & $^{64}$Ni &  $^{64}$Ni & Very small amounts in n-rich layers close to the mass cut. \\
			Cu & \textbf{$^{63}$Cu} & \textbf{$^{63}$Ni}, $^{63}$Cu, \textbf{$^{63}$Zn} & As $^{63}$Ni close to mass cut, as $^{63}$Zn in the complete Si burning zone. \\
			   & \textbf{$^{65}$Cu} & $^{65}$Ni, $^{65}$Cu, $^{65}$Zn, $^{65}$Ga  & As $^{65}$Ga in complete Si burning zone, as others close to mass cut.\\
			Zn & \textbf{$^{64}$Zn} &  $^{64}$Zn, $^{64}$Ga, $^{64}$Ge & All made during complete Si burning, high $^{64}$Zn in $Y_e$ > 0.5 zone.\\
			   & \textbf{$^{66}$Zn} & $^{64}$Ni, \textbf{$^{66}$Zn},  \textbf{$^{66}$Ge} & As $^{66}$Zn close to mass cut, as $^{66}$Ge in the complete Si burning zone. \\
			   & $^{67}$Zn &  $^{67}$Ni, \textbf{$^{67}$Cu}, $^{67}$Zn, $^{67}$Ga &  Mostly small amounts, as $^{67}$Cu in n-rich layers close to the mass cut.    \\
			   &  $^{68}$Zn & \textbf{$^{68}$Ni}, \textbf{$^{68}$Zn}, $^{68}$Ge, $^{68}$Se &  Mostly small amounts in n-rich layers close to the mass cut.    \\
			   & $^{70}$Zn &  $^{70}$Zn & Small amounts in n-rich layers close to the mass cut. \\
			\tableline
		\end{tabular}
	\end{center}
	\tablecomments{The stable isotopes (column 2) that make strong contributions to the yield of each element are given in bold typeface. For each stable isotope, the parent isotopes (column 3) that typically make strong contributions, if identifiable, are also given in bold typeface.}
\end{table*}

\begin{table*}
	\begin{center}
		\caption{Explosive synthesis of intermediate mass elements (Si-Ca)}
        	\label{tab:intermed_synth}
		\begin{tabular}{lllll}
			\tableline \tableline
			Element & Stable Isotopes & Made As  & Production \\
			\tableline
            Si & \textbf{$^{28}$Si} & $^{28}$Si & Depleted during Si burning, made via O burning.\\
               &  $^{29}$Si & $^{29}$Al, \textbf{$^{29}$Si}  & Primarily made as itself during explosive O burning. \\
               & $^{30}$Si &\textbf{ $^{30}$Si}, $^{30}$P & Primarily made as itself during explosive O burning.\\
			P  & $^{31}$P  & \textbf{$^{31}$P}, $^{31}$Si& Mostly made as itself during explosive O burning.\\
            S  & \textbf{$^{32}$S} & $^{32}$Si, $^{32}$P, \textbf{$^{32}$S} & Made as itself via O burning\\
               & $^{33}{\rm S}$ & $^{33}{\rm P}$, \textbf{$^{33}$S} & Mostly made as itself during explosive O burning. \\
               & $^{34}{\rm S}$ & $^{34}{\rm S}$ & Made as itself during explosive O burning.  \\
               & $^{36}{\rm S}$ & \textbf{$^{36}$S}, $^{36}$Cl & Made during explosive O burning.   \\
			Cl & \textbf{$^{35}$Cl} & $^{35}$P, $^{35}$S, \textbf{$^{35}$Cl} & Primarily made as itself during explosive O burning.\\
               & $^{37}{\rm Cl}$  & $^{37}$Cl, \textbf{$^{37}$Ar} &  Mostly made as $^{37}$Ar during explosive O burning.\\ 
			Ar & \textbf{$^{36}$Ar} & $^{36}{\rm Ar}$ & Made via O burning. \\
               & $^{38}$Ar & \textbf{$^{38}$Ar}, $^{38}$K & Primarily made as $^{38}$Ar during explosive O burning. \\
               & $^{40}$Ar & $^{40}$Cl, \textbf{$^{40}$Ar}, $^{40}$K & Made as $^{40}$Ar and $^{40}$Cl during explosive O burning. \\
            K  & $^{39}$K & $^{39}$K  &  Explosive O burning and complete Si burning.\\
               & $^{41}$K & $^{41}$Ca & Explosive O burning and complete Si burning. \\
            Ca & \textbf{$^{40}$Ca$^{*}$ }& $^{40}$Ca &  Primarily during incomplete Si burning. \\
               & $^{42}$Ca  & $^{42}$Ca & In $Y_e$ > 0.5 layers and explosive oxygen burning. \\
               & $^{43}$Ca & $^{43}$Ar, $^{43}$K, $^{43}$Ca \textbf{$^{43}$Sc} & Mostly as $^{43}$Sc in $Y_e$ > 0.5 layers. \\
               & $^{44}$Ca & $^{44}$Ar, $^{44}$Ca, \textbf{$^{44}$Ti} & Mainly as $^{44}$Ti in Si burning zones. \\
               & $^{46}$Ca$^{*}$ & $^{46}$Ca & Small amounts in n-rich layers close to the mass cut.\\
            \tableline
		\end{tabular}
	\end{center}
	\tablecomments{* indicates meta-stable isotopes. The stable isotopes (column 2) that make strong contributions to the yield of each element are given in bold typeface. For each stable isotope, the parent isotopes (column 3) that typically make strong contributions, if identifiable, are also given in bold typeface.
}
\end{table*}

%%%%%%%%%%%%%%%%%%%%%%%%%%%%%%%%%%%%%%%%%%%%%%%%%%%
\subsection{Two representative models: s16.0 and s21.0}
\label{subsec:s16_s21}

Next, we discuss two representative models, s16.0 and s21.0. 
They are models from the WHW02 series with relatively low and relatively high compactness. The pre-explosion properties of both models are listed in Table~\ref{tab:prop_s16_s21}, along with the explosion energies and remnant masses (``mass cut'') obtained in the hydrodynamic simulation. The corresponding yields of the isotopes $^{56,57,58}$Ni and $^{44}$Ti are also listed.

\begin{table*}
	\begin{center}
		\caption{Pre-explosion and explosion properties for models s16.0 and s21.0, with corresponding yields of selected Ni and Ti isotopes
        	\label{tab:prop_s16_s21}
        }
		\begin{tabular}{llllllllllllll}
			\tableline \tableline
			M$_{\rm ZAMS}$& {$\xi_{2.0}$} & M$_{\rm prog}$ & M$_{\rm Fe}$ & M$_{\rm CO}$ & M$_{\rm He}$ & M$_{\rm env}$ & E$_{\rm expl}$ & M$_{\rm remn}$ &  Layer & $^{56}{\rm Ni}$ & $^{57}{\rm Ni}$ & $^{58}{\rm Ni}$ & $^{44}{\rm Ti}$\\
			(M$_{\odot}$) & (-)   & (M$_{\odot}$)  & (M$_{\odot}$)     & (M$_{\odot}$)     & (M$_{\odot}$) & (M$_{\odot}$)  & (B)  & (M$_{\odot}$) & (-) & (M$_{\odot}$) & (M$_{\odot}$) & (M$_{\odot}$) & (M$_{\odot}$)\\
			\tableline
			16.0 & 0.23 & 13.25 & 1.36 & 3.34 & 4.49 & 8.75 & 1.24 & 1.54 & Si-O & 7.54$\times 10^{-2}$ & 2.97$\times 10^{-3}$ & 7.41$\times 10^{-3}$ & 3.26$\times 10^{-5}$\\
			21.0 & 0.46 & 12.99 & 1.46 & 5.55 & 6.79 & 6.21 & 1.46 & 1.82 & Si-O & 1.35$\times 10^{-1}$ & 2.84$\times 10^{-3}$ & 1.85$\times 10^{-3}$  & 6.14$\times 10^{-5}$ \\
			\tableline
		\end{tabular}
	\end{center}
	\tablecomments{The table columns are: zero age main sequence (ZAMS) mass, compactness at bounce, total mass at collapse, mass of the iron core, carbon-oxygen core, and helium core, mass of the hydrogen-rich envelope, explosion energy ($1\; {\rm B}= 1\;{\rm Bethe} = 10^{51}~\rm erg$) and remnant mass. The column after the remnant mass indicates which layer in the pre-explosion structure this corresponds to. The last four columns give the nucleosynthesis yields of observable isotopes of nickel and titanium. 
	}
\end{table*}

The pre-explosion composition of the inner layers of the s16.0 and s21.0 models is shown in the top panels of Figure~\ref{fig:profiles_prog_endab}, where the mass fractions of the most abundant alpha nuclei are plotted as a function of enclosed mass. We can see that, outside of the iron core, the composition of the material is dominated by mostly silicon  (and sulfur) and then mostly oxygen further outside.
The isotopic profiles outside of the mass cut obtained after explosive burning are shown in the bottom panels of Figure~\ref{fig:profiles_prog_endab}, together with the peak temperature achieved in each layer. 
The composition of the innermost ejected layers is dominated by $^{56}$Ni, with smaller amounts of $^{57}$Ni and $^{58}$Ni. The radioactive isotope $^{44}$Ti is also made in the same layers, which undergo explosive Si-burning.

For both models, the mass cut is located in the silicon-oxygen shell (indicated by the vertical dashed line).
For the s21.0 model, the mass cut is further out than for the s16.0 model. This is consistent with a broader trend seen in PUSH simulations: the remnant mass increases with increasing compactness of the pre-explosion models. Hence, in models with higher compactness, a smaller portion of the Si-shell is ejected.
However, the higher compactness model here also has a higher explosion energy. 
As a consequence, a larger portion of the Si-shell reaches high enough temperatures to undergo complete Si-burning. We refer the reader to Paper~II for a detailed discussion of the trends in remnant mass and explosion energy for the models used in this study.
We can see that the width of the complete Si-burning zone is larger in the case of s21.0. Even though a smaller part of the silicon-oxygen shell is ejected compared to s16.0, a larger portion of the ejected silicon undergoes complete Si burning. 
All of this combined results in higher yields of the symmetric isotopes $^{56}{\rm Ni}$ and $^{44}{\rm Ti}$ for s21.0 compared to s16.0.

The yields of asymmetric isotopes $^{57}{\rm Ni}$ and $^{58}{\rm Ni}$ do not exhibit a corresponding increase. This is because the production of these isotopes is also sensitive to the electron fraction in the region where they are synthesized. For s16.0, the mass cut is located 
relatively deep in the silicon-oxygen shell (at 1.54~M$_{\odot}$) where the pre-explosion $Y_{e}$ is lower ($\sim$ 0.498). This favors the production of neutron-rich isotopes such as $^{57}{\rm Ni}$ and $^{58}{\rm Ni}$. Further out, the $Y_{e}$ rises at $\sim 1.93$~M$_{\odot}$, as we make the transition from the silicon-oxygen to the oxygen-neon shell. The mass fraction of $^{56}{\rm Fe}$, which makes the composition slightly neutron rich, drops precipitously at this location. For s21.0, the mass cut lies towards the outer edge of the silicon-oxygen shell (at 1.82~M$_{\odot}$) in a region with higher $Y_{e}$ ($\sim$ 0.499). In this case, the transition from low $Y_{e}$ values has already occurred at $\sim 1.73$~M$_{\odot}$ (i.e.\ inside the Si-burning region). A clear effect of the $Y_{e}$ value on both $^{57}{\rm Ni}$ and $^{58}{\rm Ni}$ can also be seen in the case of s21.0, where we note that the production of both isotopes is enhanced in the small region of minimum $Y_{e}$. Nucleosynthesis of these asymmetric isotopes thus depends sensitively on the properties of the pre-explosion model such as composition, electron fraction, and also on the mass cut.

\begin{figure*}  %%%  
\begin{center}
	\begin{tabular}{cc} 
\includegraphics[width=0.49\textwidth]{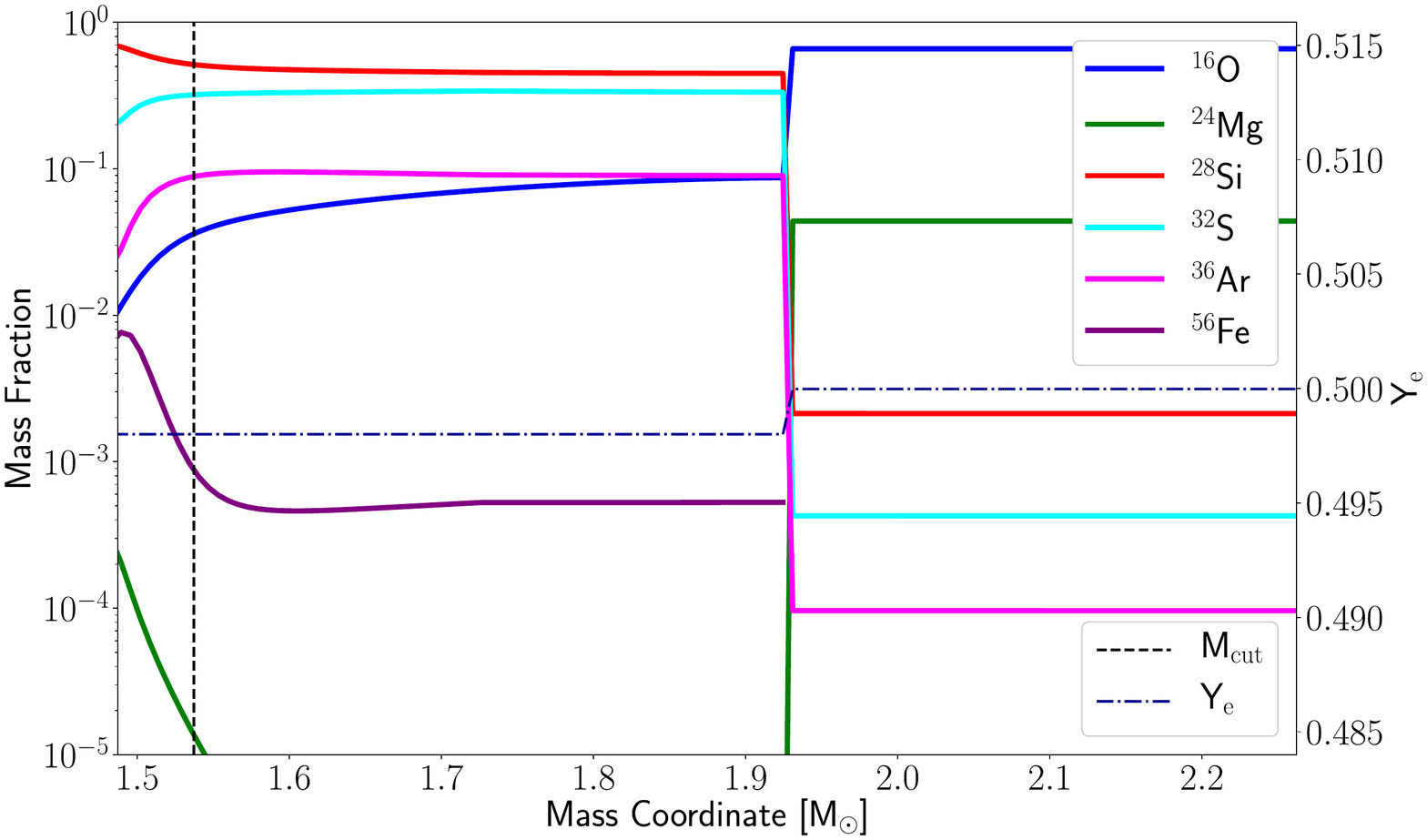}
\includegraphics[width=0.49\textwidth]{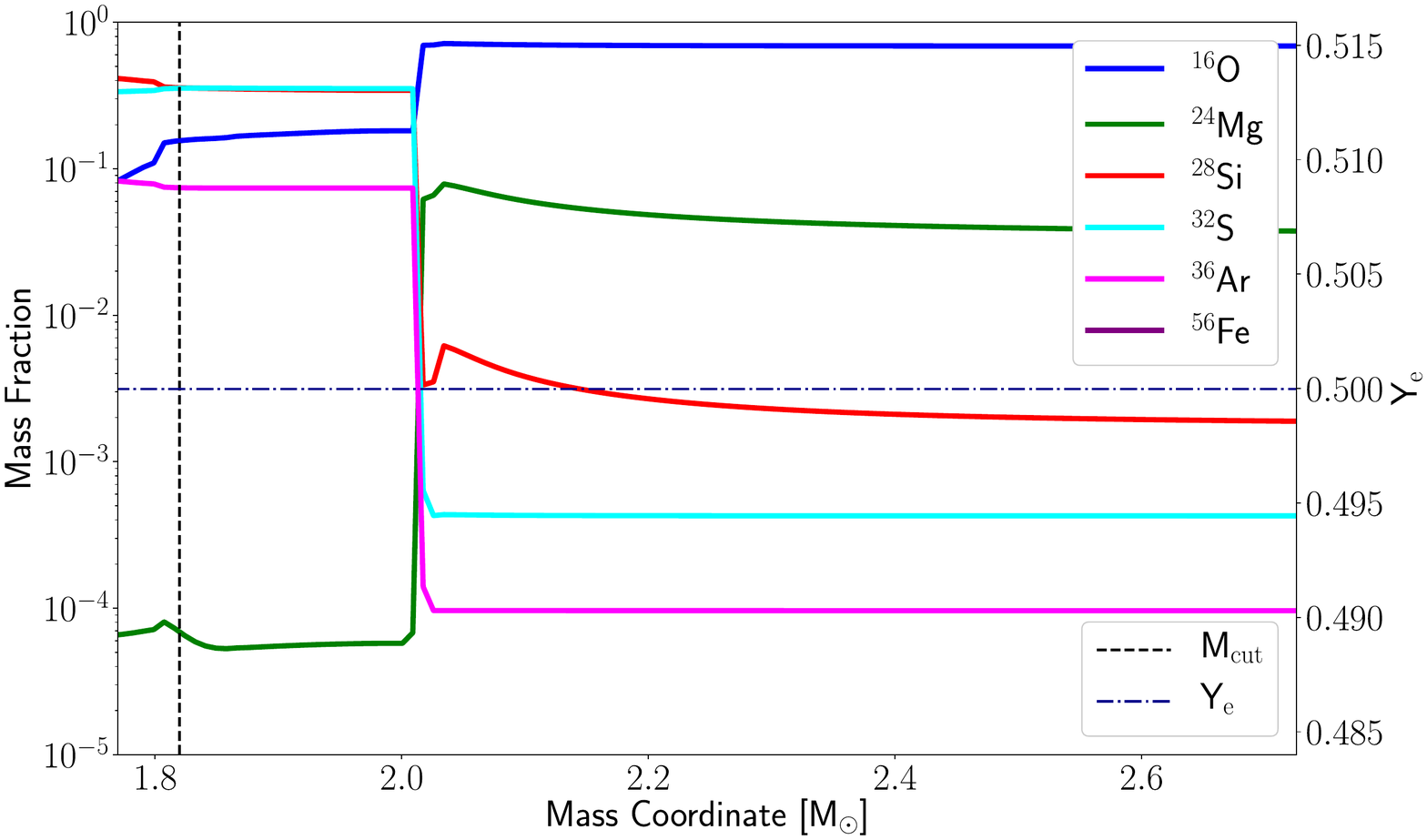} \\
\includegraphics[width=0.49\textwidth]{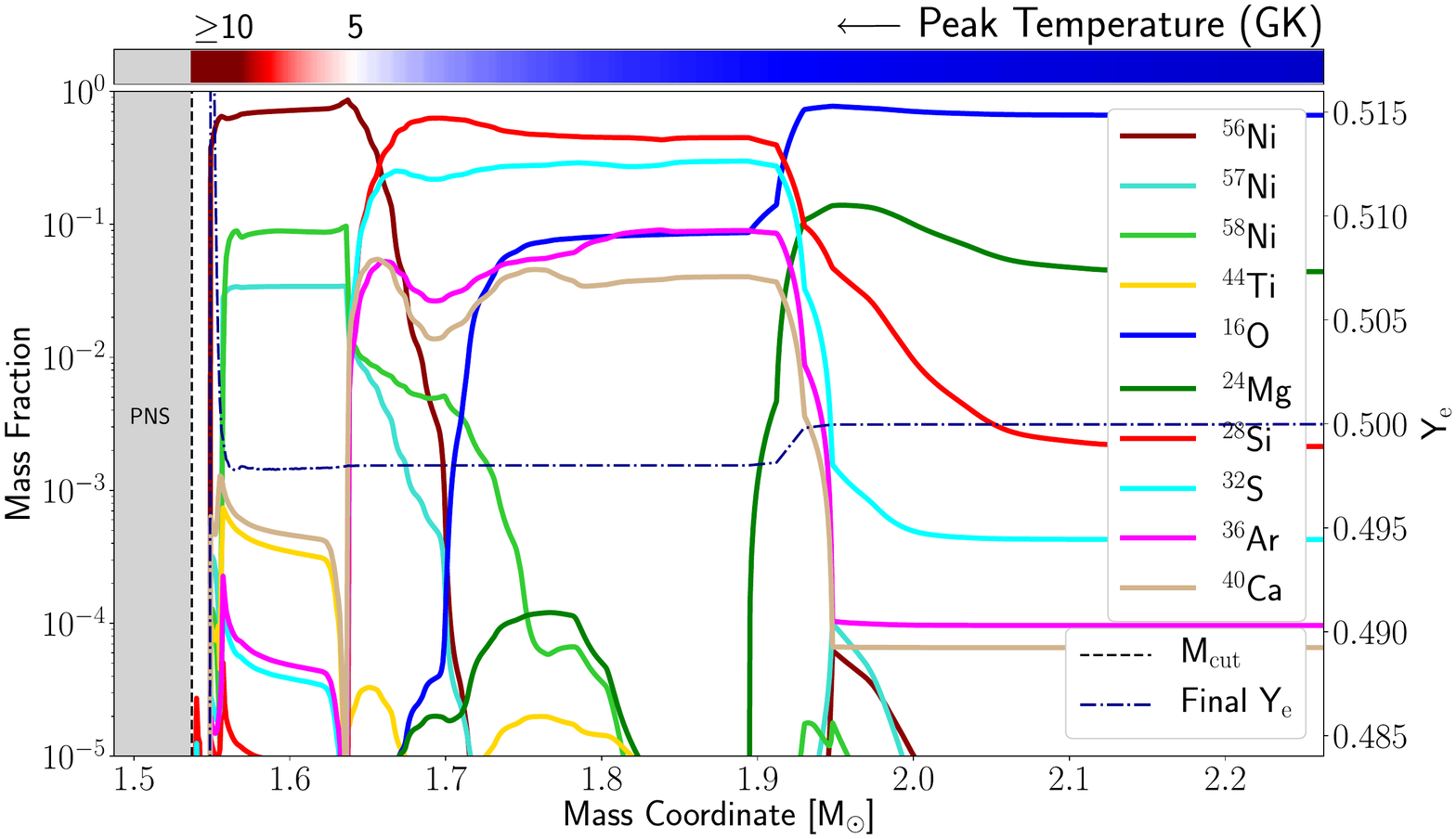}
\includegraphics[width=0.49\textwidth]{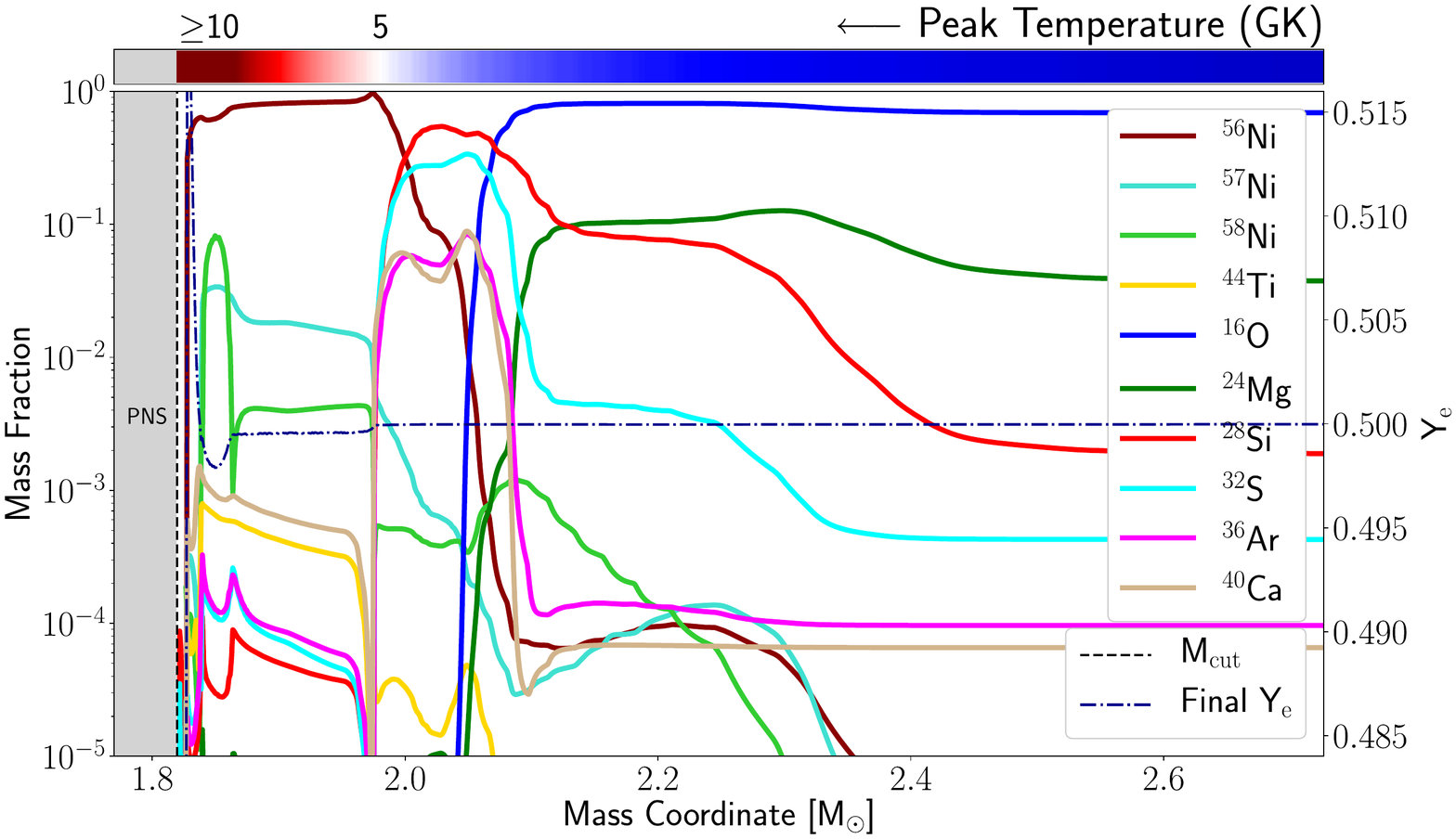}	
\end{tabular}
\caption{Top panels: composition of the inner layers of the s16.0 (left) and s21.0 (right) models prior to explosion. 
Bottom panels: peak temperatures attained in each layer (color bar) and final composition after explosive nucleosynthesis for the s16.0 (left) and s21.0 (right) models. The vertical dashed lines indicate the position of the mass cut. The pre-explosion and final electron fraction profiles are shown as dot-dashed lines. Note: In the top right panel, the Si-O shell of the progenitor has nearly equal mass fractions of Si and S (and hence the two lines overlap between 1.8 and 2.0~M$_{\odot}$).
	\label{fig:profiles_prog_endab}
    } 
\end{center}
\end{figure*}

The complete yields for both models, together with the calibration model for SN~1987A (s18.8) are given in Appendix~\ref{appx:tables} (Table~\ref{tab:finabs02}).
Figure~\ref{fig:A_Y_s16_s21} shows the integrated abundances of all layers that reach a peak temperature above 1.75~GK as function of mass number (corresponding to 0.73~M$_{\odot}$ of material for s16.0 (red circles) and 0.91~M$_{\odot}$ for s21.0 (blue stars)). Even though the $^{56}$Ni yields are different by a factor of 2 (see Table~\ref{tab:prop_s16_s21}), overall the yields look quite similar between the two models. The main differences are for the intermediate mass elements ($24 \leqslant A \leqslant 44 $) and to a lesser degree for elements beyond iron ($A \approx 90$). The intermediate mass elements are mostly synthesized during incomplete Si-burning. As seen in Figure~\ref{fig:Tmax_Rhomax}, while the conditions for complete Si-burning are very similar between the two models (even though the corresponding amount of mass is different), the conditions for incomplete Si-burning are somewhat different. The lower compactness model s16.0 experiences slightly higher densities for the same peak temperature. 
More importantly, in the s21.0 model, virtually the entire Si-shell undergoes complete, explosive Si-burning. Hence, any elements between Si and Fe have to be synthesized from intermediate mass elements such as Ne and Mg. For s16.0, the situation is quite different. Less than half of the Si-shell undergoes complete Si-burning, leaving plenty of material with mass number around $A \sim 28$ that can be processed to up to $A \sim 44$ during explosive burning.
For both models, there are some late-time neutron-rich ejecta ($Y_e$ $\sim$ 0.42), which dominate the production of elements beyond iron (up to $A \sim 140$) via a weak r-process. These abundances are quite uncertain and detailed comparison between the two models remains beyond the scope of this work.

\begin{figure}  %%% 
\includegraphics[width=0.49\textwidth]{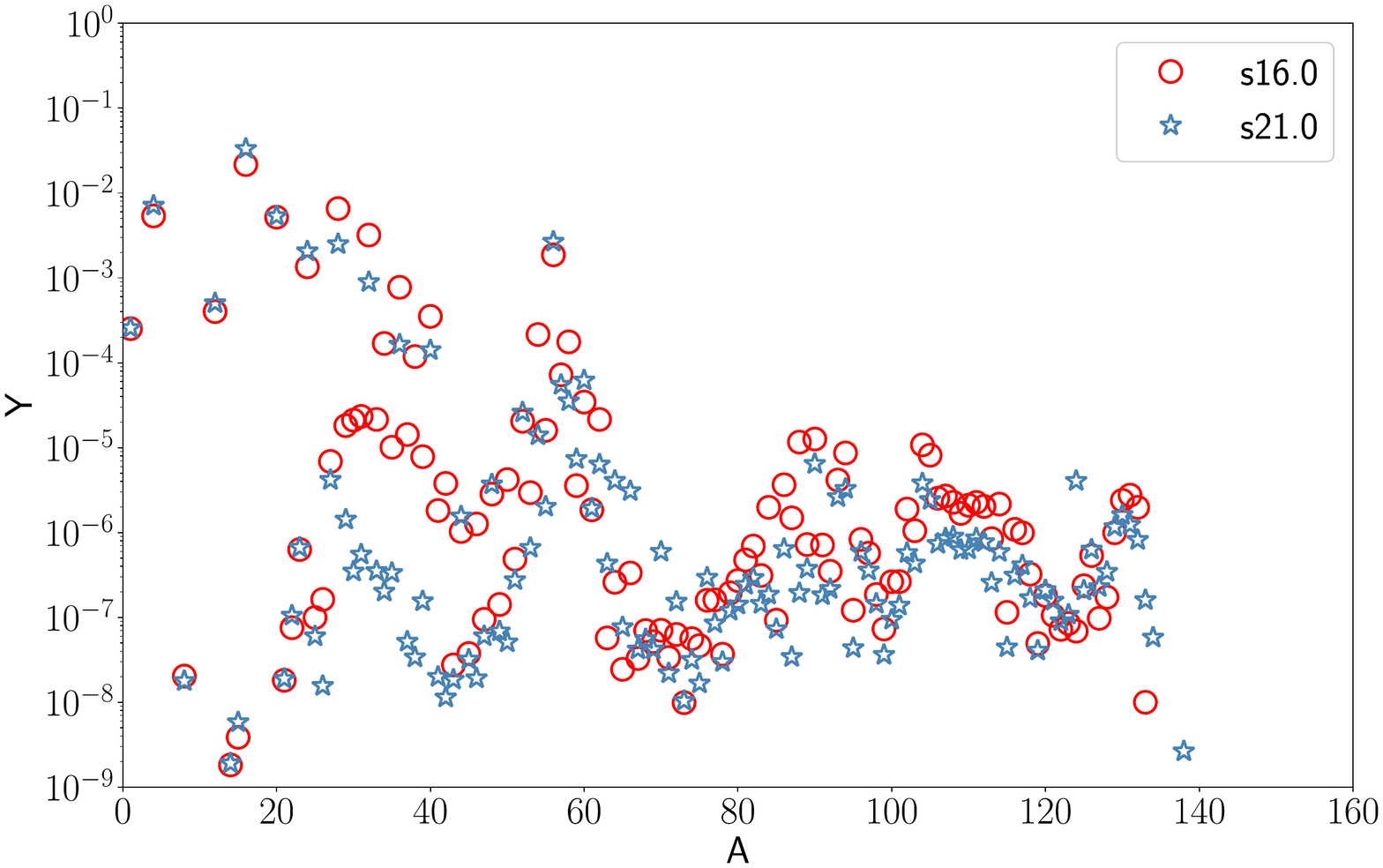}
\caption{Mass-integrated final abundances $Y$ as function of mass number $A$ for models s16.0 (red circles) and s21.0 (blue stars). Only material with peak temperatures of $T_{\mathrm{peak}} \geqslant 1.75$~GK is included.
	\label{fig:A_Y_s16_s21}
}
\end{figure}

\begin{figure}  %%%  
\includegraphics[width=0.49\textwidth]{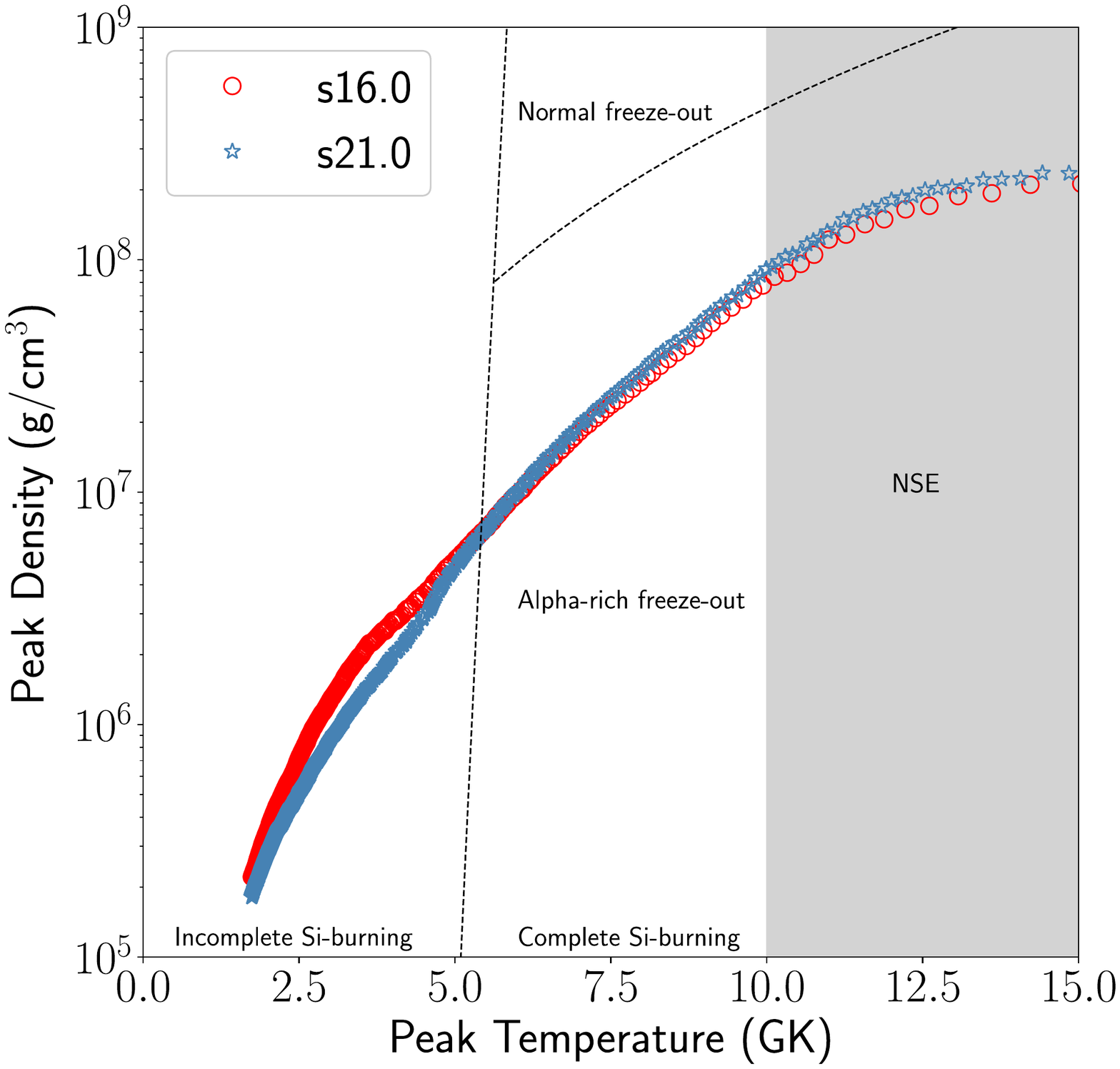}
\caption{Peak density and peak temperature for the two representative models s16.0 (blue open stars) and s21.0 (red open circles) for layers which reach $T_{\mathrm{peak}} \geqslant 1.75$~GK.
	\label{fig:Tmax_Rhomax}
}
\end{figure}

%%%%%%%%%%%%%%%%%%%%%%%%%%%%%%%%%%%%%%%%%%%%%%%%%%%
\subsection{Trends with progenitor and explosion properties} 
\label{subsec:comptrend}

We now investigate the combined nucleosynthesis yields from all models included in this study to identify any common behaviors that exist across the spectrum. By including two sets of solar-metallicity pre-explosion models, we span nucleosynthesis uncertainties introduced by differences in the pre-explosion stellar evolution.

Examining the predicted yields as a function of ZAMS mass of the pre-explosion model does not show any apparent correlations. However, strong correlations do emerge for some of the isotopes as well as elements when we consider the yields as a function of the compactness of the pre-explosion models.
The $^{56}{\rm Ni}$ yields, for example, increase linearly with compactness, as shown in Figure~\ref{fig:nitiye}. Other symmetric isotopes such as $^{44}$Ti also show this linear correlation. The production of these isotopes depends strongly on the explosion energy and on how much of the silicon-oxygen shell gets explosively processed and ejected.
The neutron-rich isotopes $^{57}$Ni and $^{58}$Ni show no obvious single trend, but we do find two or more branches that appear to independently correlate with compactness. The color-coding in Figure~\ref{fig:nitiye} indicates the electron fraction of the region where each isotope is made. 
This reveals that, for some asymmetric isotopes, even small changes in the local $Y_e$ can far exceed the impact of other relevant factors (such as the compactness, ZAMS mass, or location of the mass cut) in setting the final yields of these isotopes.
We have explicitly tested this for a few models by making small changes ($\sim$0.001) to the initial $Y_e$ during post-processing, while keeping all other quantities unchanged. We found that the final mass fractions of certain isotopes were affected by up to an order of magnitude in the layers where the $Y_e$ had been altered. In general, low $Y_e$ values favor the production of neutron-rich isotopes such as $^{57}$Ni and $^{58}$Ni, which can explain the trends we see here. \cite{nakamura1999}, \cite{Umeda2002} and \cite{2005umeda} have also previously found a strong dependence of the cobalt and manganese abundances on the $Y_e$ of the pre-explosion model. In a future work, we will investigate a few important isotopes and the detailed reaction pathways that result in such highly $Y_e$-sensitive yields. 

An additional observation we have made is that models from the WHW02 series typically fall into one of two bins, high or low $Y_e$, while the models from the WH07 series have an intermediate $Y_e$ value. One major difference between the two series is that the WH07 series employed a much larger network during stellar evolution than the WHW02 series (other differences may also be present).
A Table with ZAMS mass and yields of $^{56}$Ni, $^{57}$Ni, $^{58}$Ni, and $^{44}$Ti (as shown in Figure~\ref{fig:nitiye}) for all exploding models can be found in Appendix~\ref{appx:key_iso_all} (Table ~\ref{tab:nitisample}).

We have already described (see Section~\ref{subsec:s16_s21}) how the production of the asymmetric Ni-isotopes is sensitive to the location of the mass cut and the $Y_e$-profile relative to the pre-explosion model structure. Here, we can see the same behavior across the entire sample of models. It is clearly visible how small changes in the $Y_e$ of the material can change the relative and absolute amounts of isotopes by an order of magnitude or more. This emphasizes the need for accurate $Y_e$ values already from the pre-explosion modeling.

We also see similar correlations in the final elemental yields of different iron-group elements with compactness, shown in Figure~\ref{fig:nimncomp_ye}. The trends for these elements can be understood from the isotopes as which they are made. Titanium and chromium are predominantly made as symmetric isotopes $^{48}$Cr and $^{52}$Fe respectively. Their yields are thus linearly correlated with compactness. Most of the manganese is made as asymmetric $^{55}$Co in the incomplete Si-burning region. Nickel yields depend on both $^{58}$Ni and $^{60}$Cu apart from the symmetric isotope $^{60}$Zn. As a result, both these elements show trends that are differentiated based on the $Y_e$ in the region where they are predominantly produced.

The intermediate mass elements have significant contributions from the hydrostatic burning prior to the explosion. We do not find any clear trends for these elements with compactness or explosion energy.

\begin{figure*}  % 
\begin{center}
	\begin{tabular}{cc} 
\includegraphics[width=0.49\textwidth]{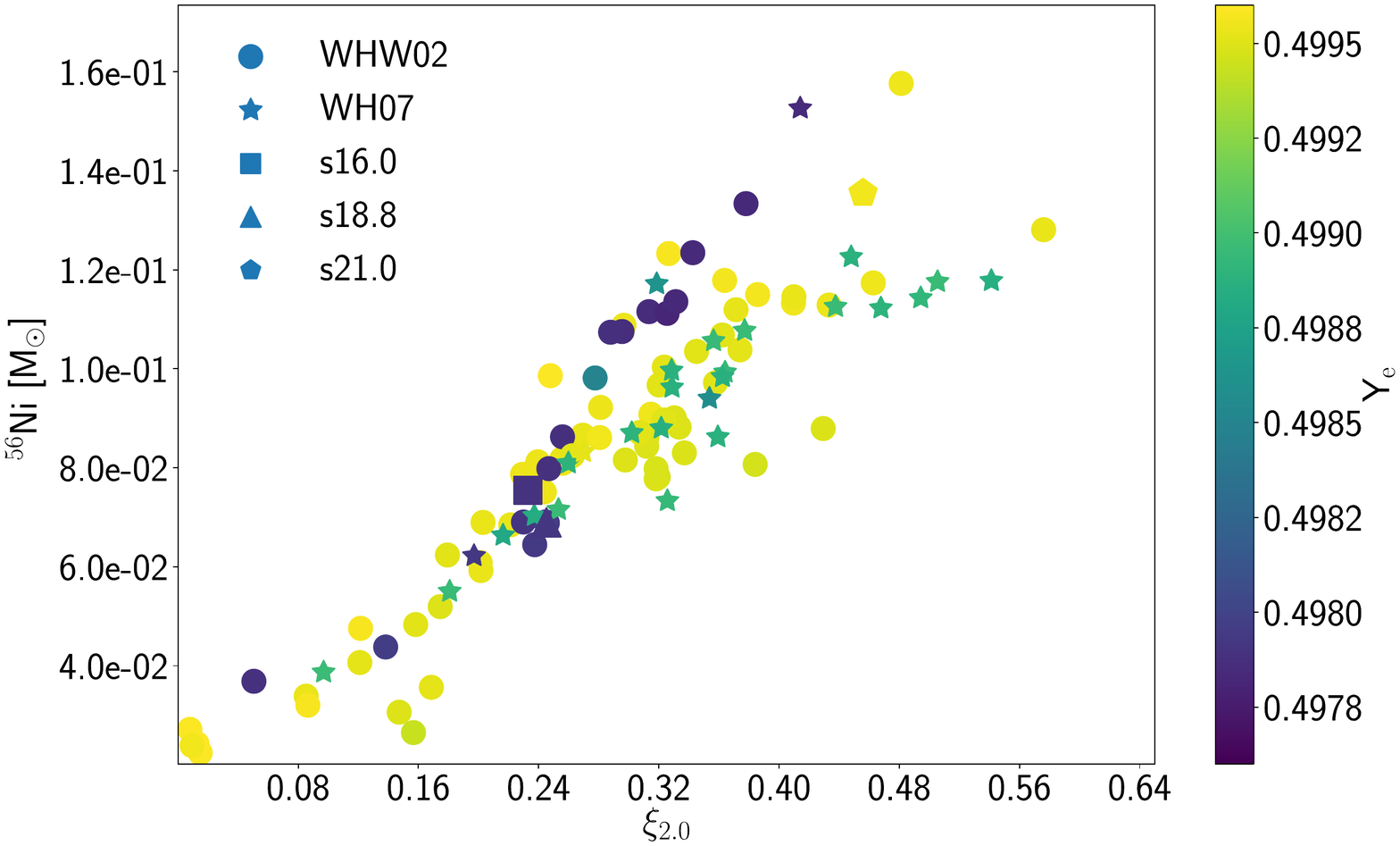}
\includegraphics[width=0.49\textwidth]{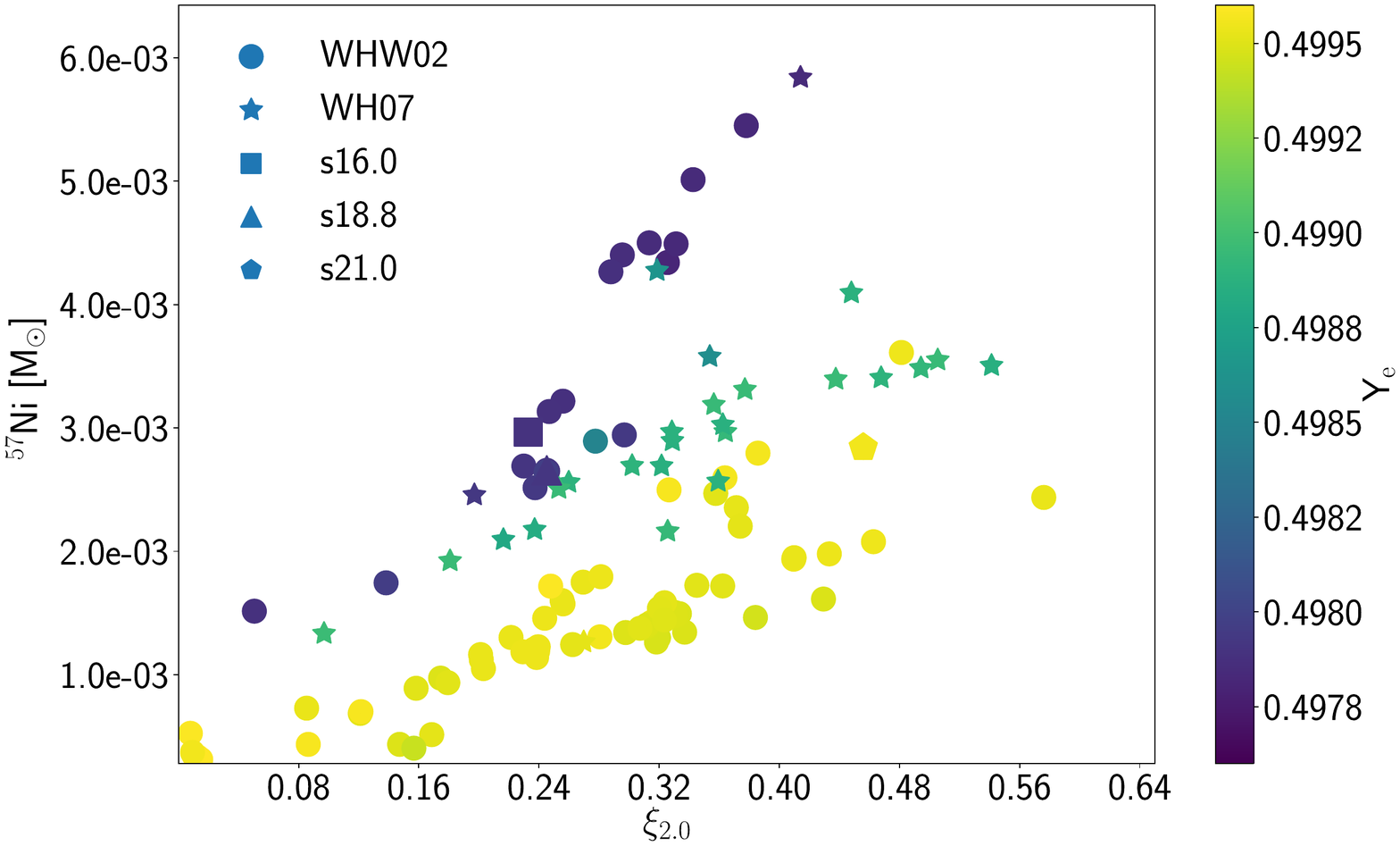}\\
\includegraphics[width=0.49\textwidth]{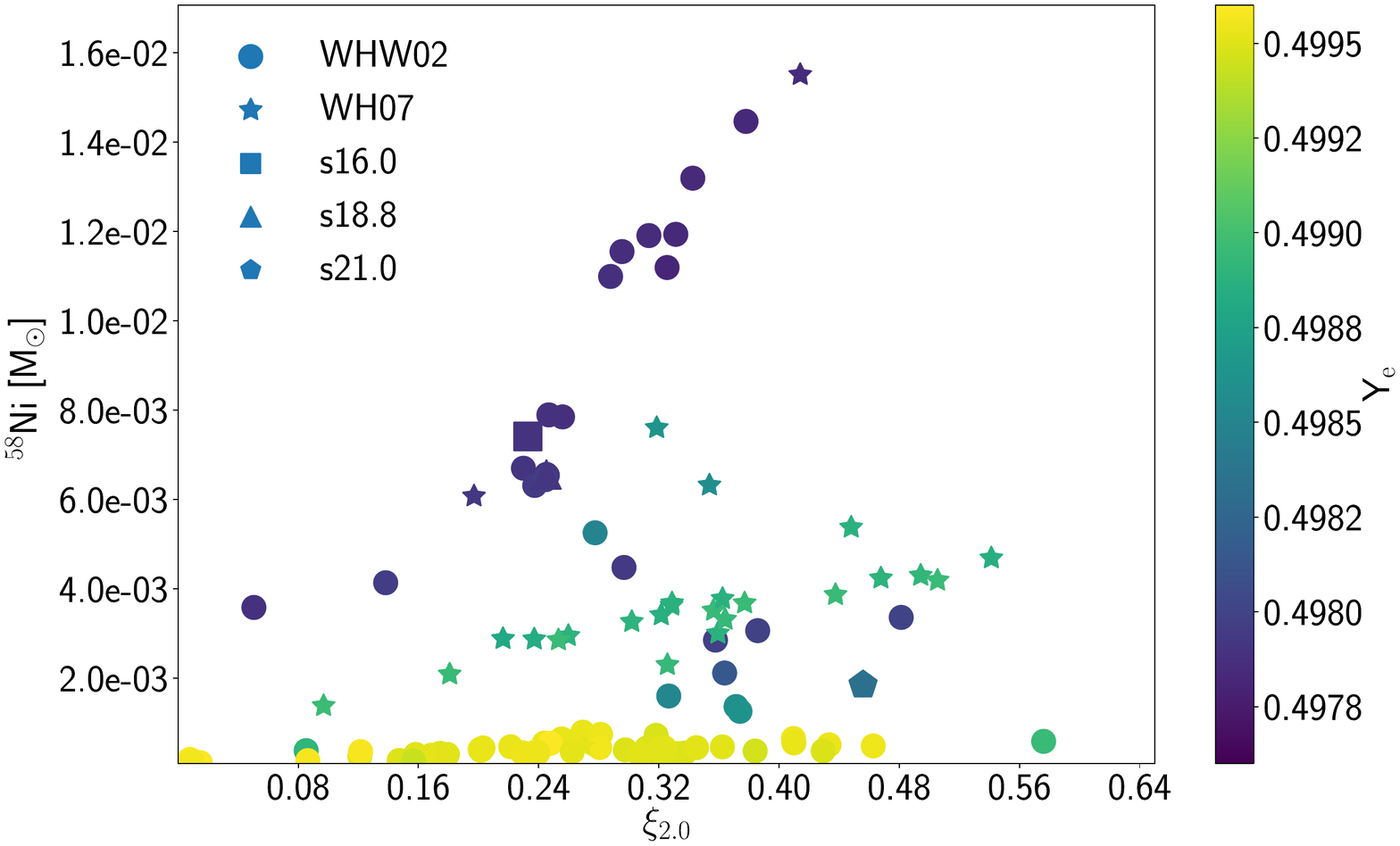}
\includegraphics[width=0.49\textwidth]{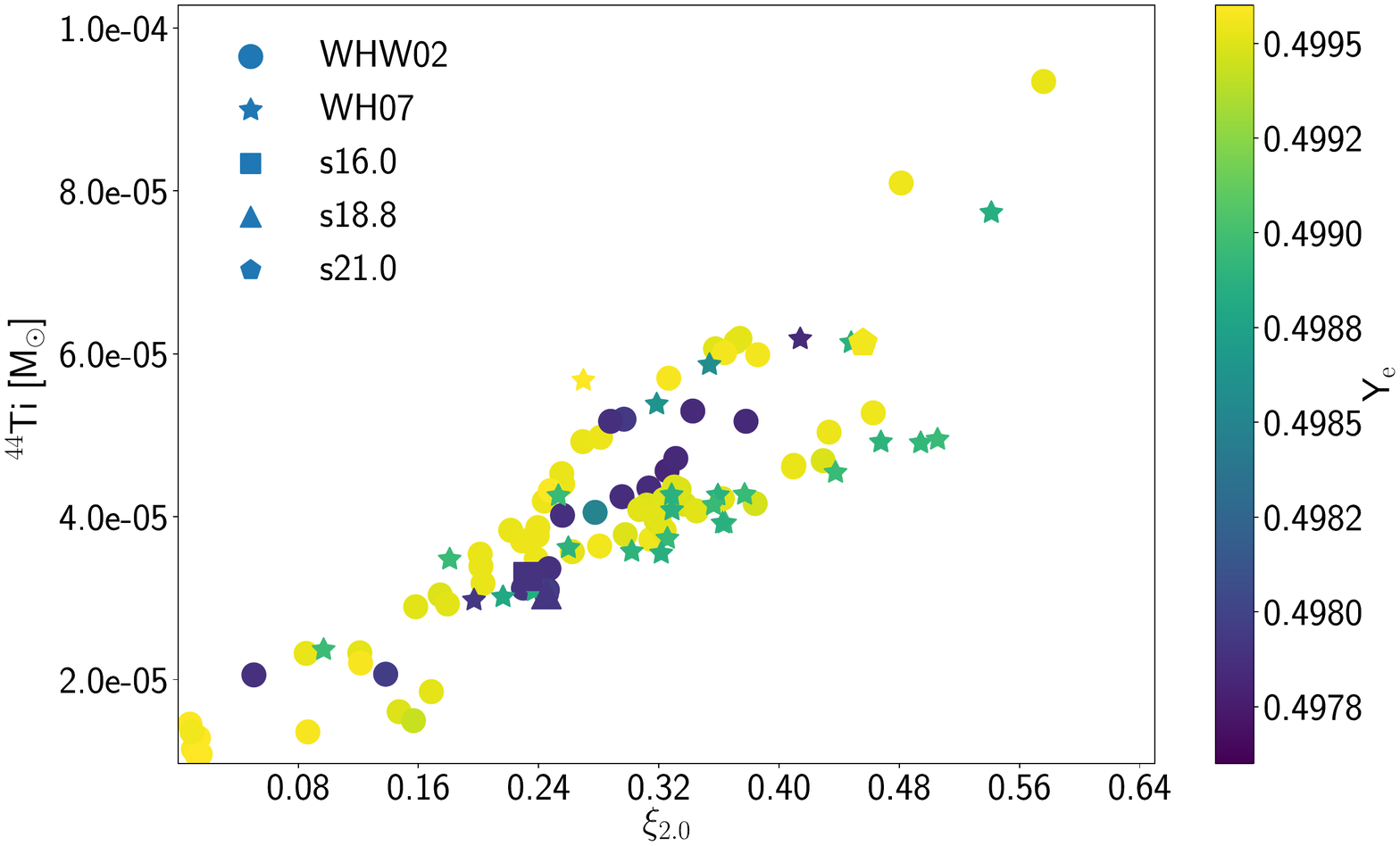}
	\end{tabular}
	 \caption{Isotopic yields of $^{56}$Ni(top left), $^{57}$Ni (top right), $^{58}$Ni (bottom left), and $^{44}$Ti (bottom right) after explosive processing as function of compactness for both sets combined (circles for WHW02; stars for WH07). The color-coding represents the average $Y_e$ value in the layers that made the highest contribution to the yield of each isotope shown. These yields can be found in Appendix \ref{appx:key_iso_all}.
		\label{fig:nitiye}
   }
\end{center}
\end{figure*}

\begin{figure*}  % 
\begin{center}
	\begin{tabular}{cc} 
\includegraphics[width=0.49\textwidth]{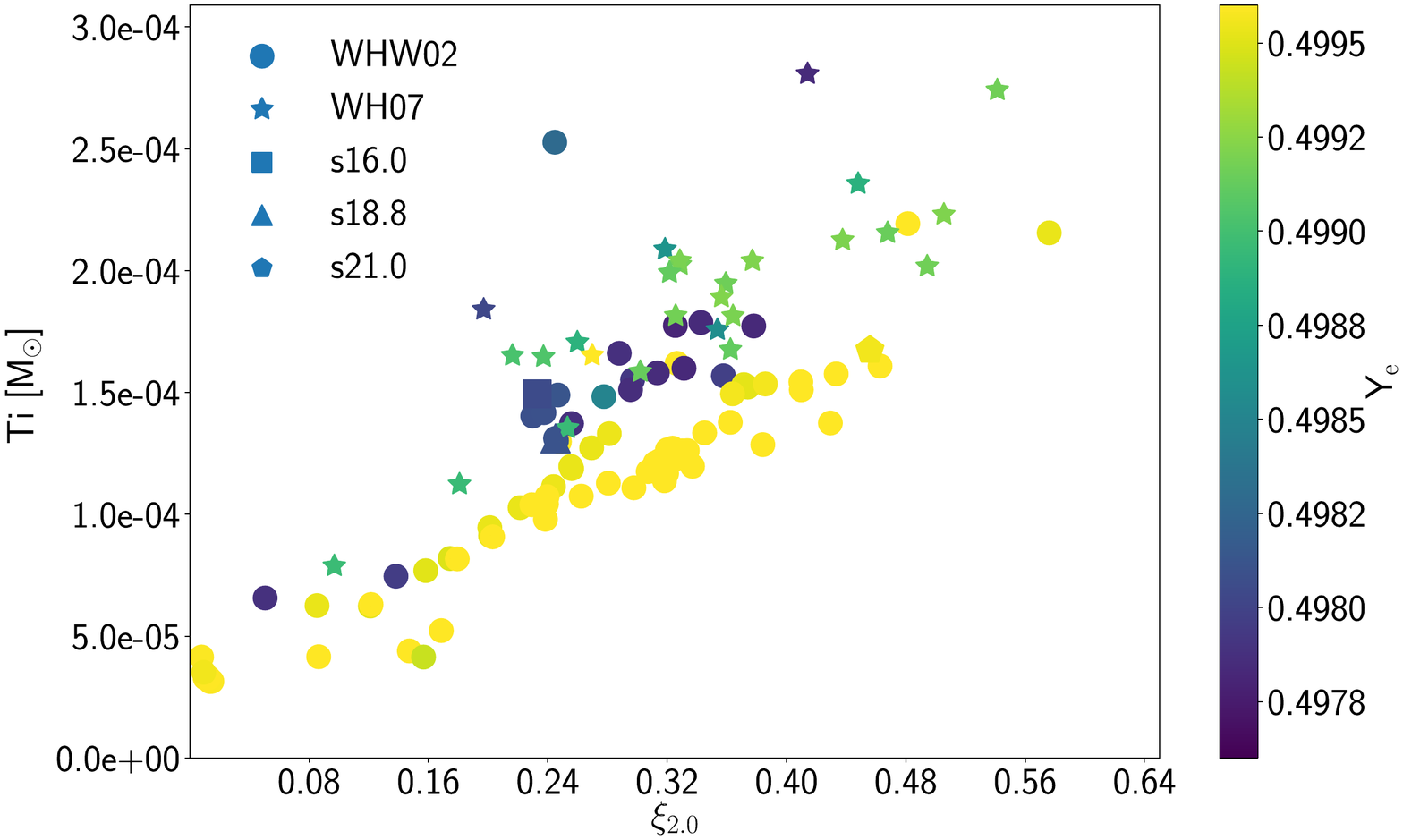} 
\includegraphics[width=0.49\textwidth]{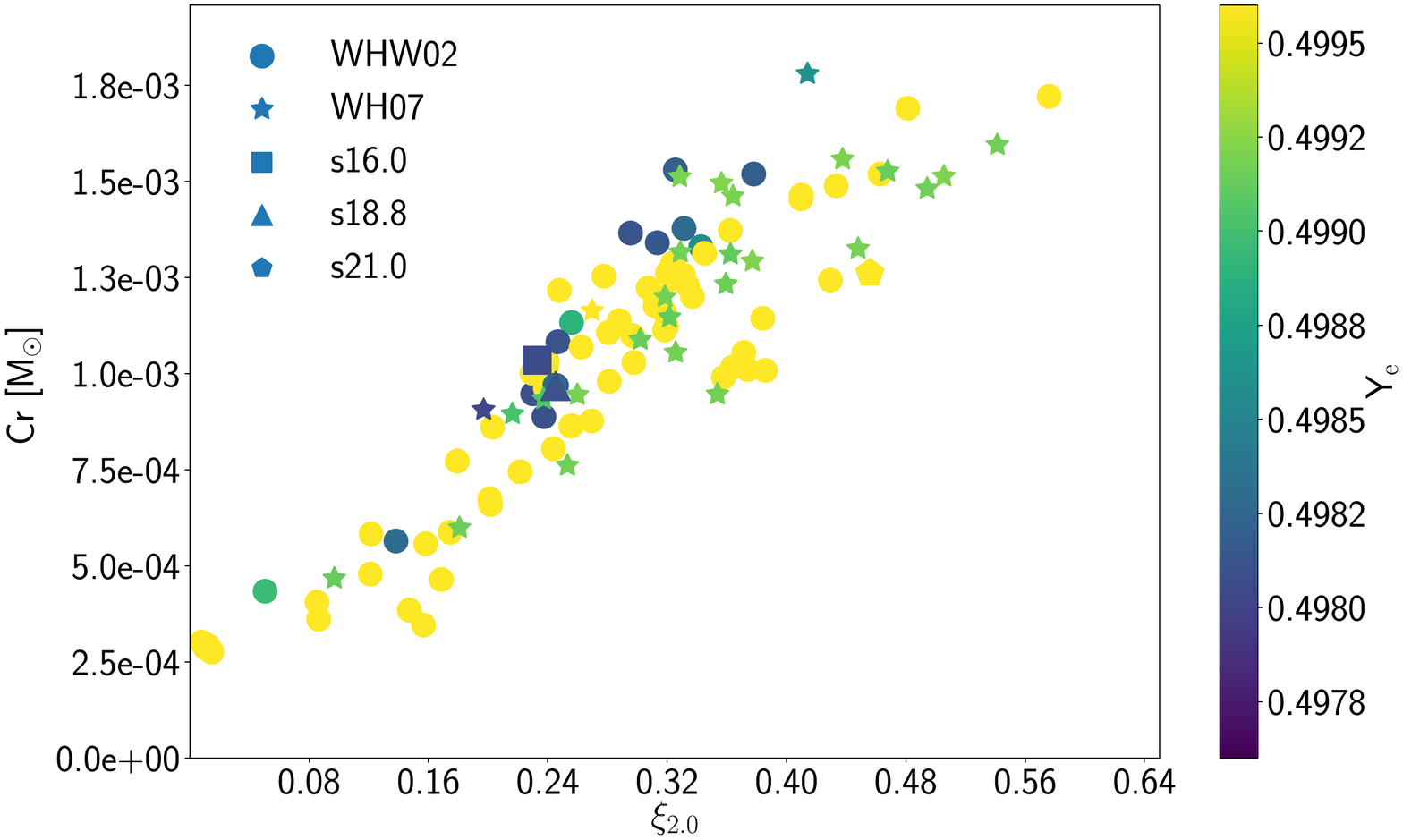}\\ 
\includegraphics[width=0.49\textwidth]{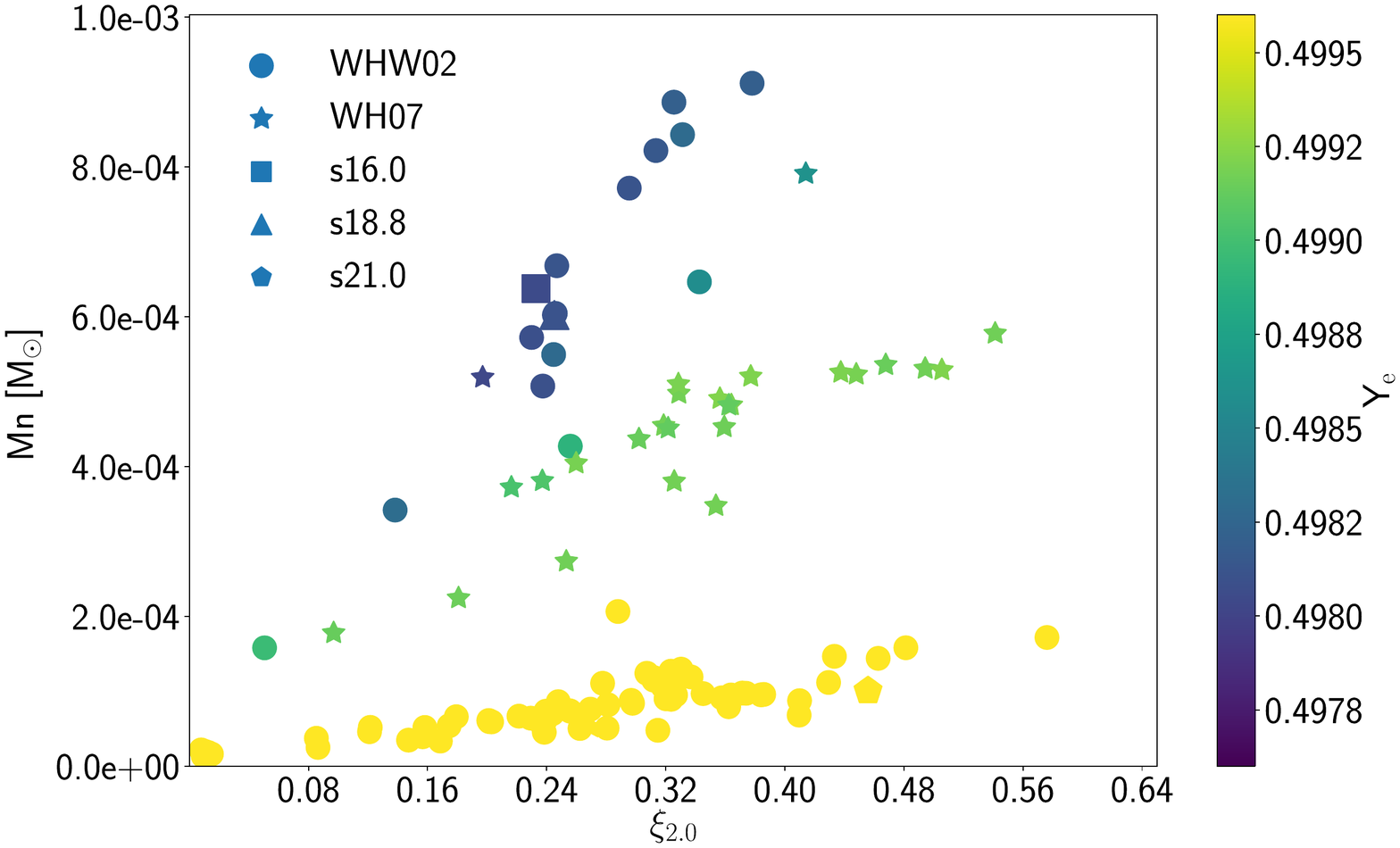} 
\includegraphics[width=0.49\textwidth]{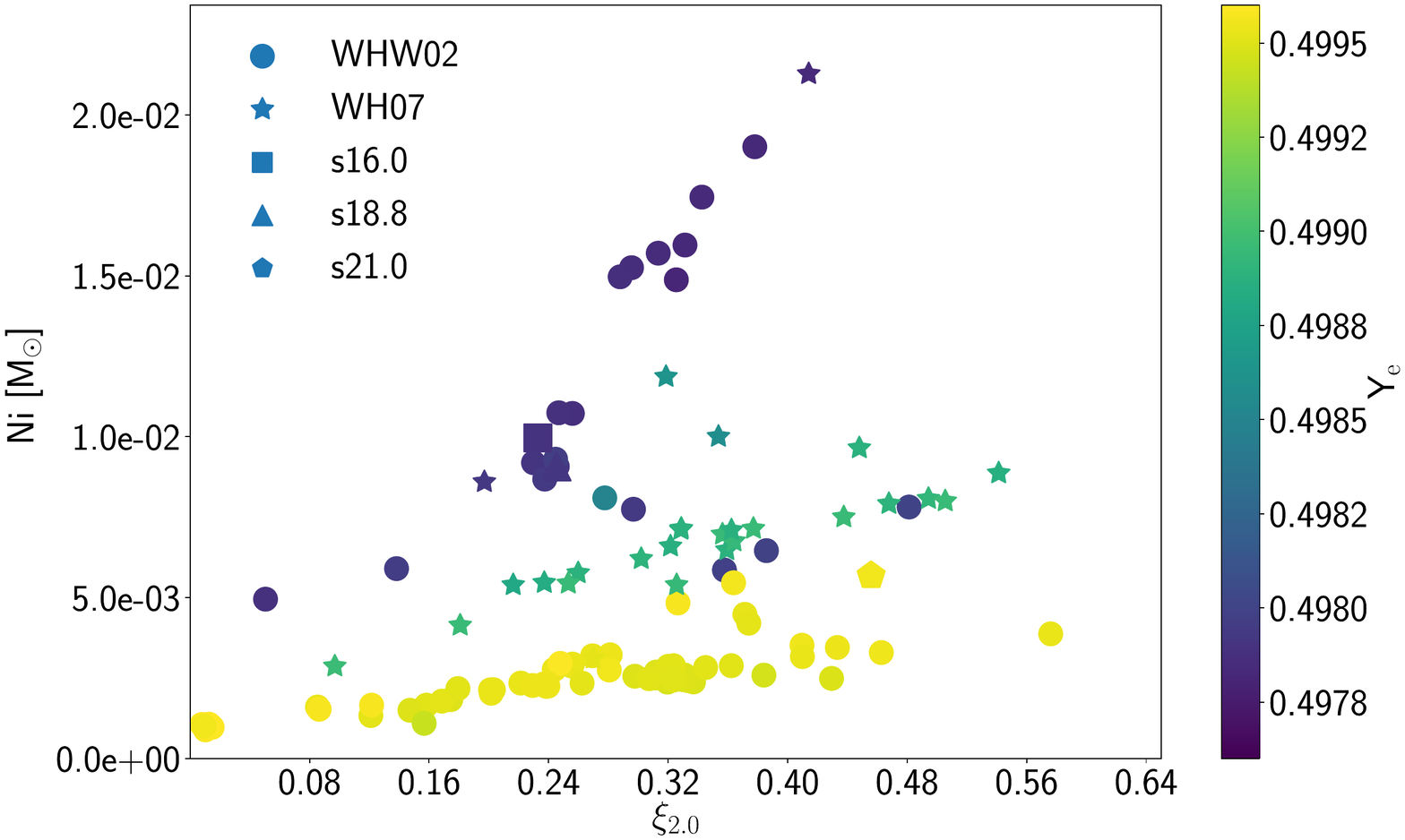}
	\end{tabular}
	\caption{Elemental yields of titanium (top left), chromium (top right), manganese (bottom left), and nickel (bottom right) after explosive processing as a function of compactness for both sets combined (circles for WHW02; stars for WH07). The color-coding represents the average $Y_e$ value in the layers that made the highest contribution to the yield of each element shown.
		\label{fig:nimncomp_ye}
    }
\end{center}
\end{figure*}

%%%%%%%%%%%%%%%%%%%%%%%%%%%%%%%%%%%%%%%%%%%%%%%%%%%
%% Section 4 "Comparison with observations"
%%%%%%%%%%%%%%%%%%%%%%%%%%%%%%%%%%%%%%%%%%%%%%%%%%%
\section{Comparison with observations} \label{sec:observations}

\subsection{The supernova landscape} 
\label{subsec:Ni-landscape}

In Paper~II, we used the pre-explosion model s18.8 to calibrate the PUSH method and reproduce the observed properties of SN~1987A, including the $^{56}$Ni yield. Here, we have predicted the $^{56}$Ni yields of 110 models (excluding the calibration model s18.8) based on PUSH-induced explosions. 
In Figure~\ref{fig:ninomoto}, we compare our predicted $^{56}$Ni yields with yields derived from observations of core-collapse supernovae. Our calibration model s18.8 is also included in the figure.
The observed properties of these CCSNe (neutrino-driven CCSNe, excluding HNe) have been adapted from \cite{bruenn16} and \cite{nomoto13}.
We find a good match across the entire mass range between our predicted yields and the available $^{56}$Ni yields from observations. This is the case for each set of pre-explosion models individually as well as for the combined results. For ZAMS masses up to $\sim 15$~M$_{\odot}$, the amount of $^{56}$Ni rises with ZAMS mass. Above $\sim 15$~M$_{\odot}$, the amount of $^{56}$Ni ejected remains roughly constant. Note that there are no observational data for nickel above $\sim 30$~M$_{\odot}$ ZAMS mass.

In addition to the $^{56}$Ni yields, Ni/Fe ratios derived from observations of SN spectra are also available for a few core-collapse supernovae. A compilation of these ratios is available in Table~4 of \cite{2012ec}. In Figure~\ref{fig:jerkstrand}, we compare our results for models with ZAMS masses between 12.0 and 21.0~\msun to the observationally-derived ratios. The mass range of models included is chosen based on the estimated progenitor masses for the four supernovae shown. 
We find that, of the included models, the lower ZAMS masses match the reported ratios and $^{56}$Ni yields for these supernovae quite well. For SN2012ec, which has a supersolar Ni/Fe ratio and a high $^{56}$Ni yield, models s12.0 and s13.0 give the best matches.

\begin{figure}
\includegraphics[width=0.49\textwidth]{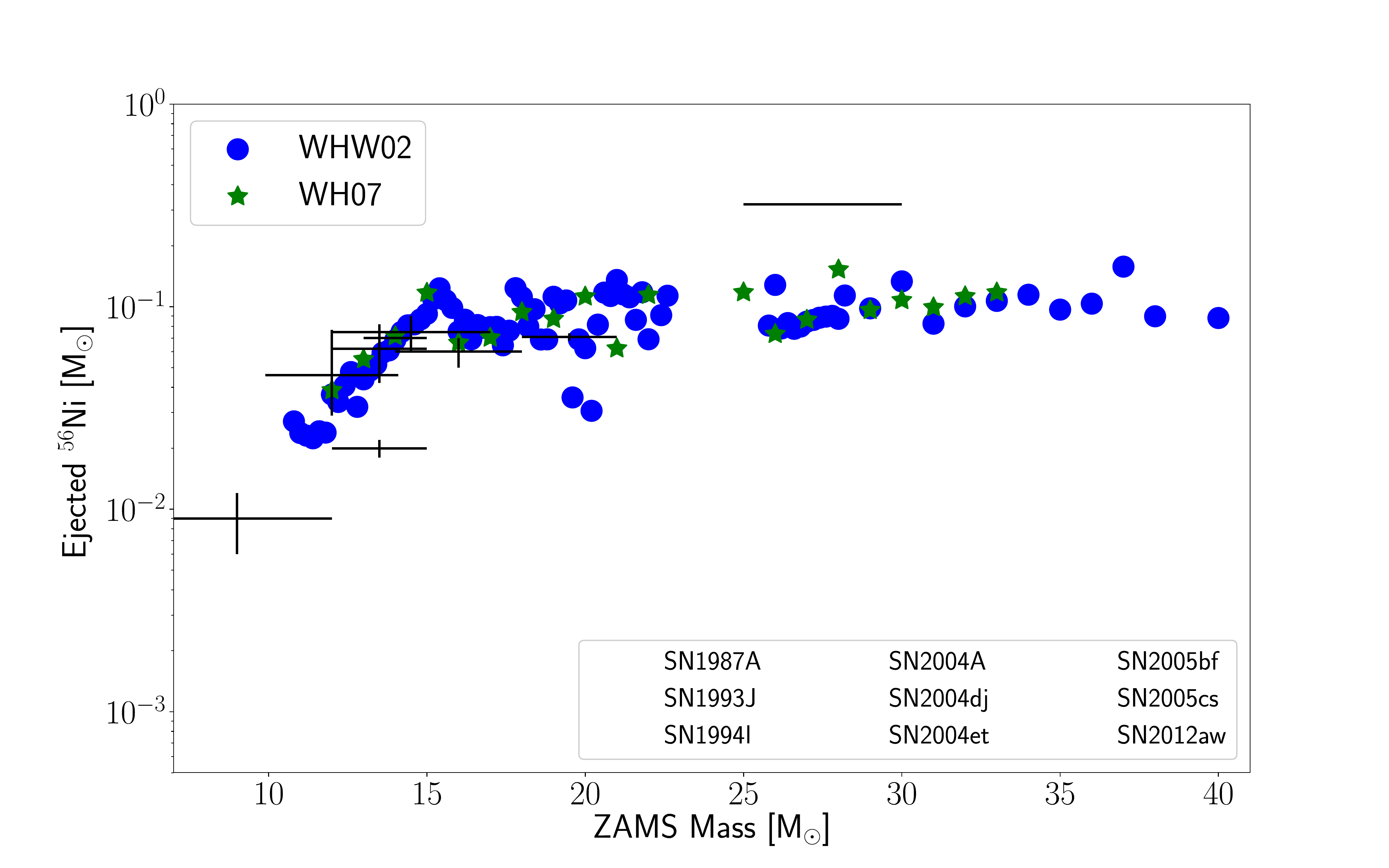}
\caption{Yields of $^{56}$Ni as function of ZAMS mass for the WHW02 (blue circles) and WH07 (green stars) models. Black error-bar crosses indicate $^{56}$Ni yields derived from observations of faint supernovae.
		\label{fig:ninomoto}
}
\end{figure}

\begin{figure}
\includegraphics[width=0.49\textwidth]{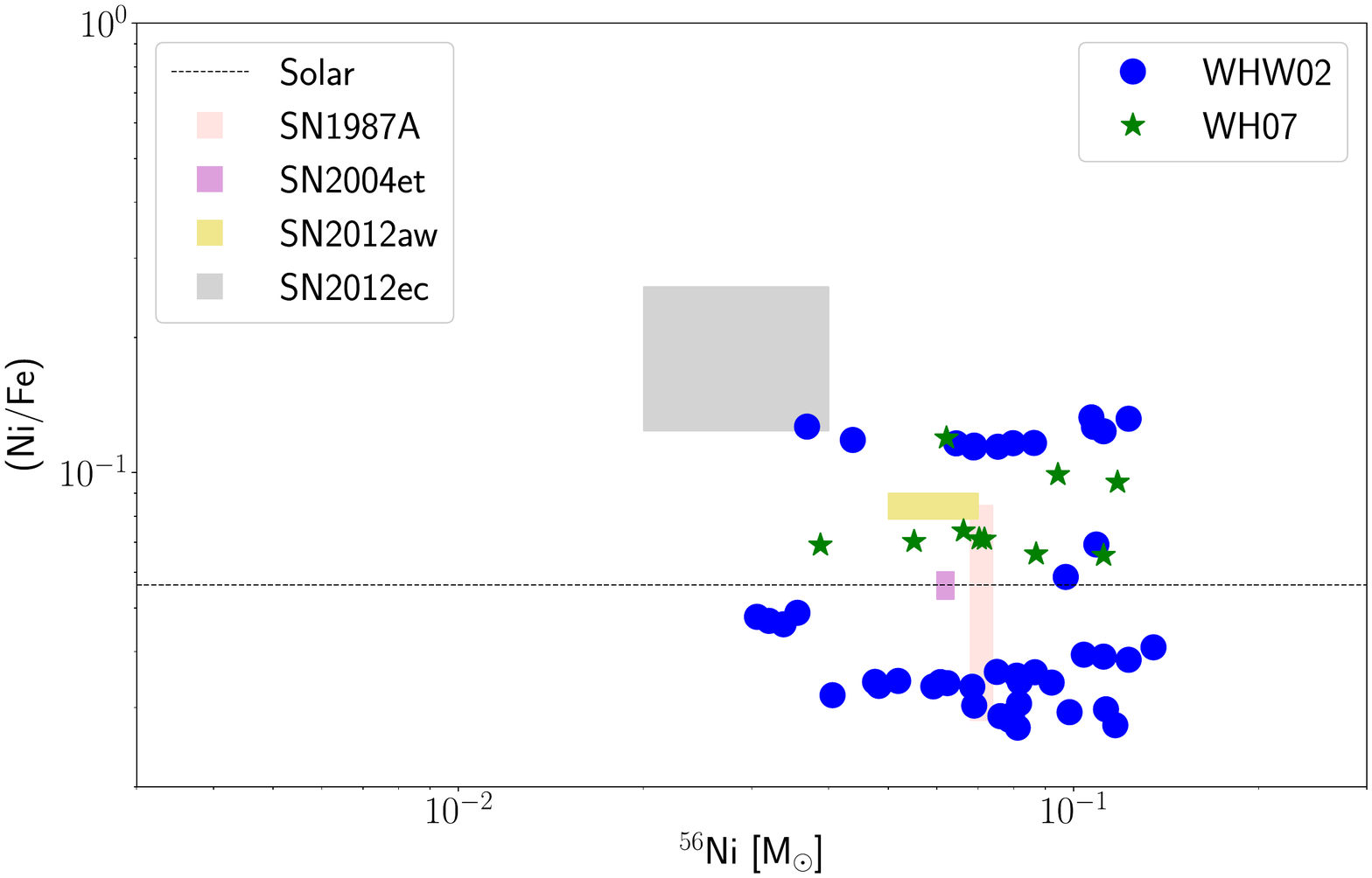}
\caption{Ni/Fe ratio as function of $^{56}$Ni ejecta for four observed SNe (see \cite{jerkstrand11} for details) and for our models of the same mass range. Note that no uncertainties in the Ni/Fe ratios were reported for SN2004et and SN2012aw, we have imbued them with a line width equivalent to the point size here for better visibility.
		\label{fig:jerkstrand}
}
\end{figure}

\subsection{Iron-group abundances in metal-poor stars} 
\label{subsec:EMPstars}

Massive stars were one of the first polluters of the universe due to their short lifetimes. Thus, the atmospheric compositions of long-lived, low-mass, metal-poor stars reflect, on average, the nucleosynthesis yields from one or few core-collapse supernovae.
As such, abundances from metal-poor stars provide an additional test for predicted CCSN yields.

There are numerous ongoing and past surveys aimed at finding metal-poors stars and determining their atmospheric abundances from spectral analyses \citep[see e.g.][]{caffau13, norris13, keller2014, li2015b, li2015a, howes15, placco15, jacobson15, hansen.hansen.ea:2016, roederer16, frebel2017, starkenburg17, andrievsky18}. 
Recently, \cite{sneden} used improved laboratory data for iron-group neutral and singly-ionized transitions to derive robust abundances in the very metal-poor main sequence turnoff star HD~84937. Figure~\ref{fig:xfeprofiles} shows the reported abundance ratios for HD~84937 along with ratios calculated from the yields predicted in this work (WHW02 models in the top panel; WH07 models in the bottom panel). The triangles indicate observational data (neutral and ionized species). Each transparent square indicates one of our models. Note that our predicted results have not been weighted with an initial mass function, rather we want to illustrate how robust (or sensitive) the synthesis of each element is.

Overall, our results are in good agreement with the observational data. We did not find a direct correlation between the ZAMS mass of the pre-explosion model and the elemental yields for elements that show model-to-model variations. Two elements, Sc and Zn, show large model-to-model variations in their yields.
Scandium has traditionally been difficult to produce in sufficient amounts in thermal bomb and piston models with a canonical explosion energy of $10^{51}$~erg. The production of Sc is enhanced when the explosion energy is higher by a factor 10, such as in hypernovae \citep{nomoto06}. Alternatively, a careful treatment of the neutrino interactions also leads to enhanced Sc production \citep{cf06a}.
In our calculations presented here, it is produced along with other iron-group elements in the innermost layers where neutrino interactions with free nucleons have created a proton-rich environment. This further underscores the importance of accounting for neutrino interactions in nucleosynthesis calculations.
Zinc shows a similar behavior to Sc. Traditional nucleosynthesis models typically underproduce Zn in explosions with energies of $\sim 10^{51}$~erg. \cite{cf06a} have found that Zn is also enhanced in CCSN simulations that follow the weak-interaction physics during collapse and explosion. In our models, we find Zn to be coproduced with other Fe-group elements in the innermost layers. Both elements are quite sensitive to the details and we find a wide range of production levels across all our exploding models, see also Figure~\ref{fig:xfeprofiles}. 
Note that the observationally derived abundances for Cu (and also for Zn) are relatively uncertain \citep{sneden}.

Another interesting effect can be seen in the typical [Mn/Fe] ratios for the two model sets. For most of the WHW02 models, the manganese yield is quite low (with some spread) whereas WH07 models consistently show a near-perfect match to the observed manganese ratio. We showed in the previous section that the Mn yield is very sensitive to the final $Y_e$ in the incomplete-silicon burning region. Since almost all WH07 models have $Y_e \sim 0.4992$ in that region, they produce higher [Mn/Fe] ratios that match the data better. The WHW02 models have either lower $Y_e$ ($\sim 0.4978$) or higher $Y_e$ ($\sim 0.4995$) in that region.

A more rigorous comparison to observed ratios requires carrying out galactic chemical evolution calculations that integrate the explosive yields from models of different ZAMS masses, also taking into account crucial factors such as the stellar initial mass function and stellar lifetimes. Such calculations are beyond the scope of this work. However, detailed tabulated yields are included with this Paper for use in chemical evolution studies.

\begin{figure}
  \includegraphics[width=0.49\textwidth]{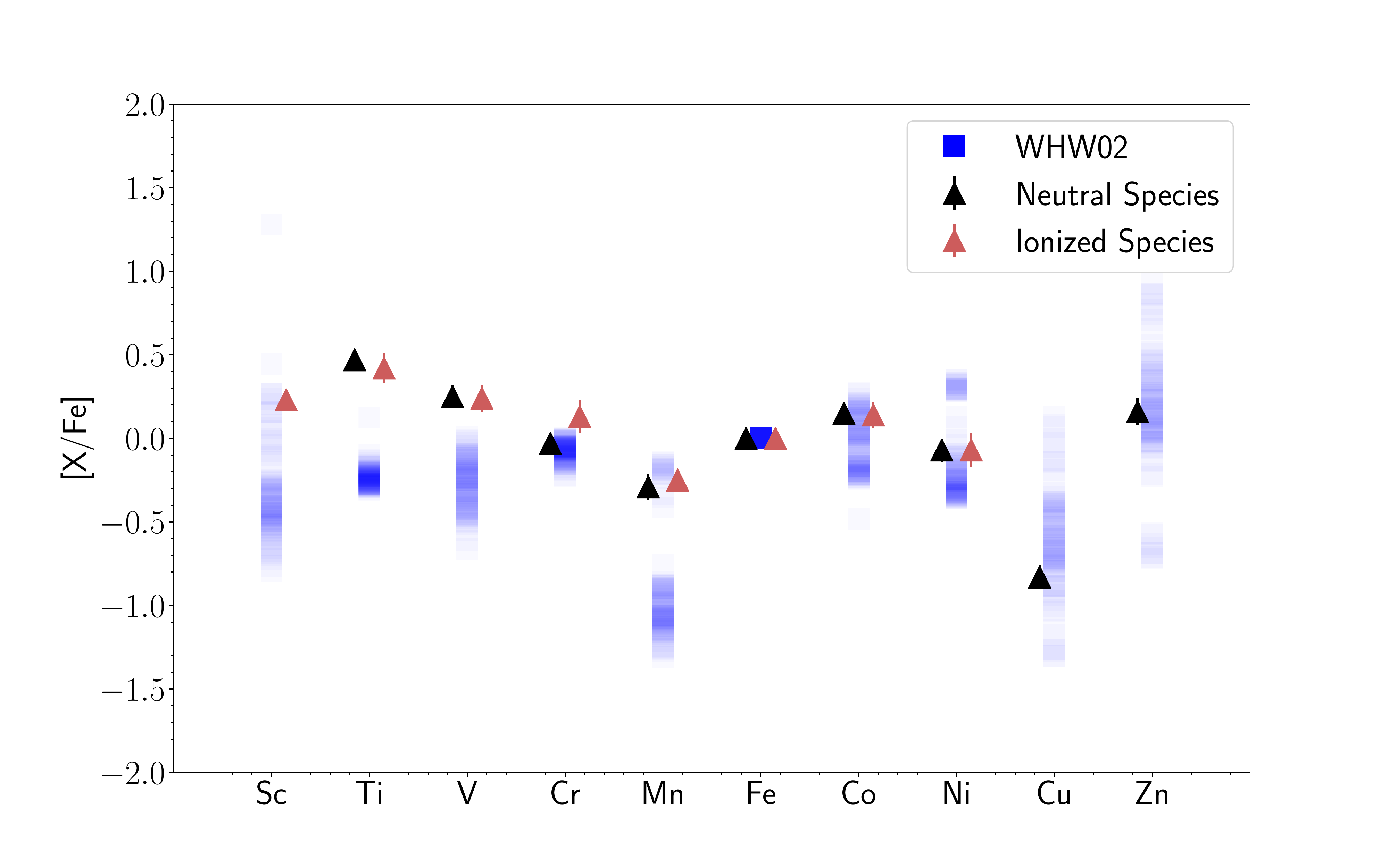}  
  \includegraphics[width=0.49\textwidth]{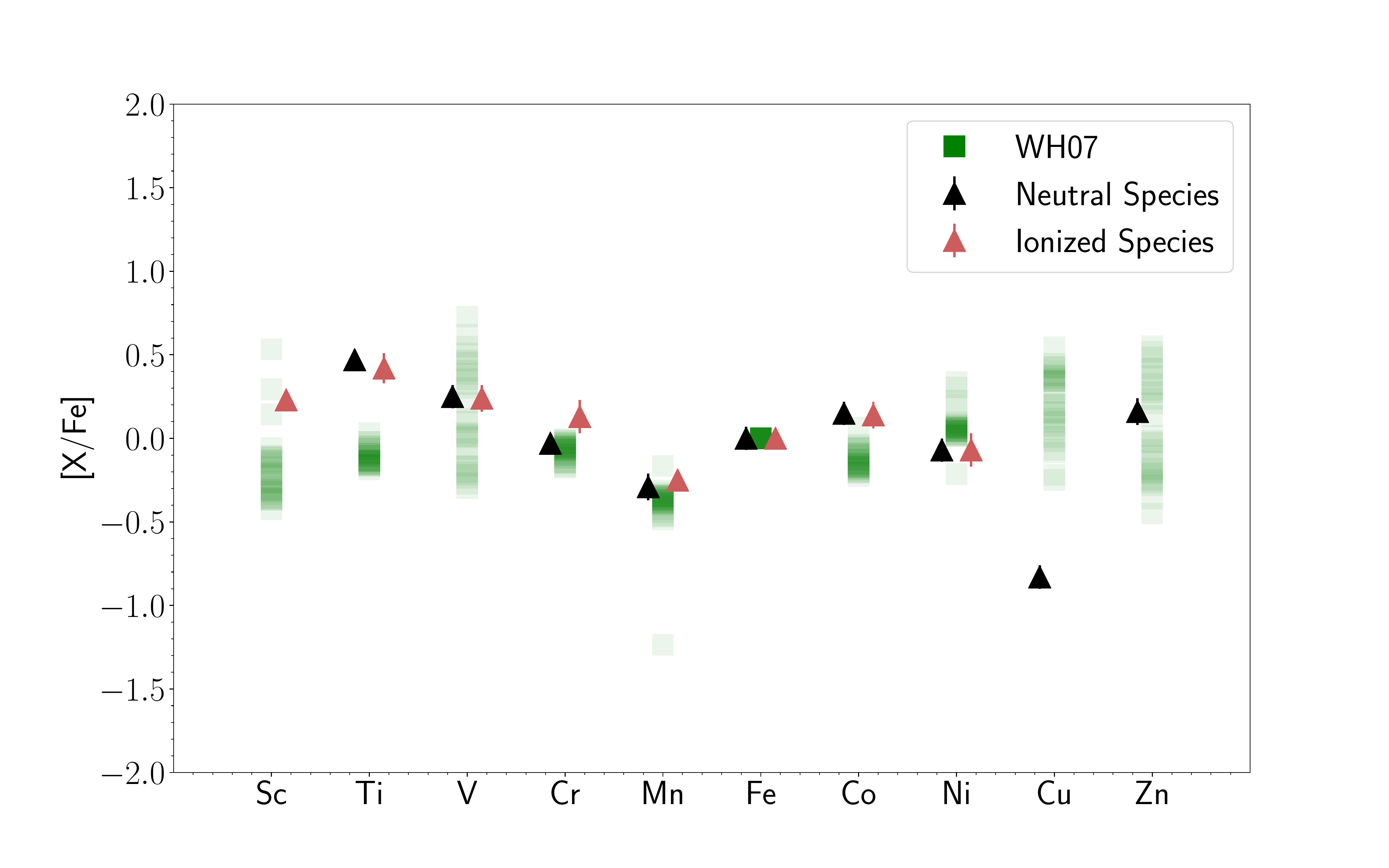}  
	\caption{Top panel: Observed abundances for HD~84937 (triangles) and our results for the WHW02 set (squares). Bottom panel: Same as the top panel but for the WH07 set. The transparency of the squares is normalized such that darker color indicates more models producing a particular element ratio and all models combined resulting in the full color as shown in the figure key. Note that the models are not weighted by an initial mass function.
		\label{fig:xfeprofiles}
		}
\end{figure}

\subsection{Alpha elements - O to Ca}
Previous sections of this paper discuss primarily the composition of the Fe-group. 
Another important product related to supernova explosions are the alpha elements, such as $^{16}$O, $^{20}$Ne, $^{24}$Mg, $^{28}$Si, $^{32}$S, $^{36}$Ar, $^{40}$Ca, $^{48}$Cr ($\rightarrow$ $^{48}$Ti), and $^{52}$Fe ($\rightarrow$ $^{52}$Cr). In low metallicity stars (with metallicities $[\mathrm{Fe}/\mathrm{H}]<-1$), the observed surface abundances generally reflect the pollution of the interstellar medium (out of which these stars formed) by earlier core-collapse supernovae. Only for $[\mathrm{Fe}/\mathrm{H}]>-1$, type Ia supernovae start to contribute, due to their delayed occurrence.

While there exists still some scatter (real and due to observational uncertainties) the averaged observed ratios of $[\alpha/\mathrm{Fe}]$ for these low metallicities are typically in the range of $+0.3$ to $+0.5$ \citep{cayrel04,nomoto13}. Thus, it is of interest whether our predictions can (on average and in their scatter) also reproduce such observed values.

In Figure~\ref{fig:ratios_alphas}, we display these values for the alpha elements O, Si, and Ca for all models in our sample (WHW02 and WH07). We see that, on average, our results are close to 0.5 for Ca and Si (alpha elements produced in explosive burning and thus directly related to the explosion mechanism). O, Ne, Mg are mostly leftovers from hydrostatic burning phases and increase with the shell size of these burning products, i.e.\ their yields increase with the ZAMS mass. In Figure~\ref{fig:ratios_alphas}, we see that $[\mathrm{O}/\mathrm{Fe}]$ is only approaching the value of 0.5 for the more massive stars above 20~M$_\odot$, while the lower mass models lead to lower values, even to values below zero. 
To interpret this result, we have to also consider that our yields are all from models with solar initial metallicity. As a function of metallicity, stellar winds increase and hence reduce the total mass of the stellar model during its evolution, including the hydrostatic yields. The observations at $[\mathrm{Fe}/\mathrm{H}]<-1$ relate to the explosive yields of stars with low metallicities (i.e. stars with much reduced mass loss). The hydrostatic yields from low-metallicity pre-explosion models should be larger and could thus resolve this apparent contradiction. In order to give an authoritative answer to this question, we will need to explore the explosions of low metallicity supernova models, which will be done in a forthcoming paper.

\begin{figure}  %%% 
\begin{center}
	\begin{tabular}{c} 
\includegraphics[width=0.49\textwidth]{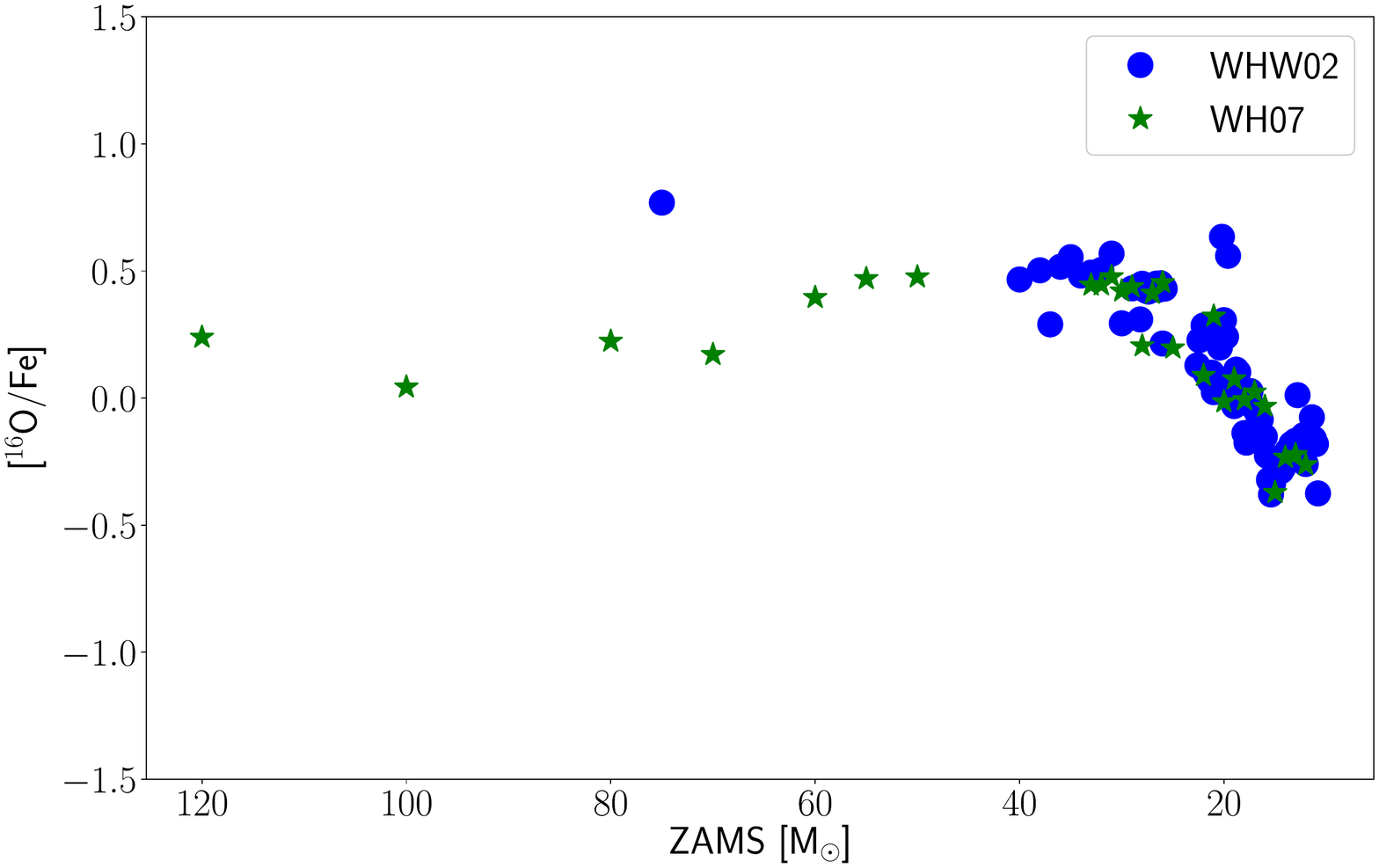} \\
\includegraphics[width=0.49\textwidth]{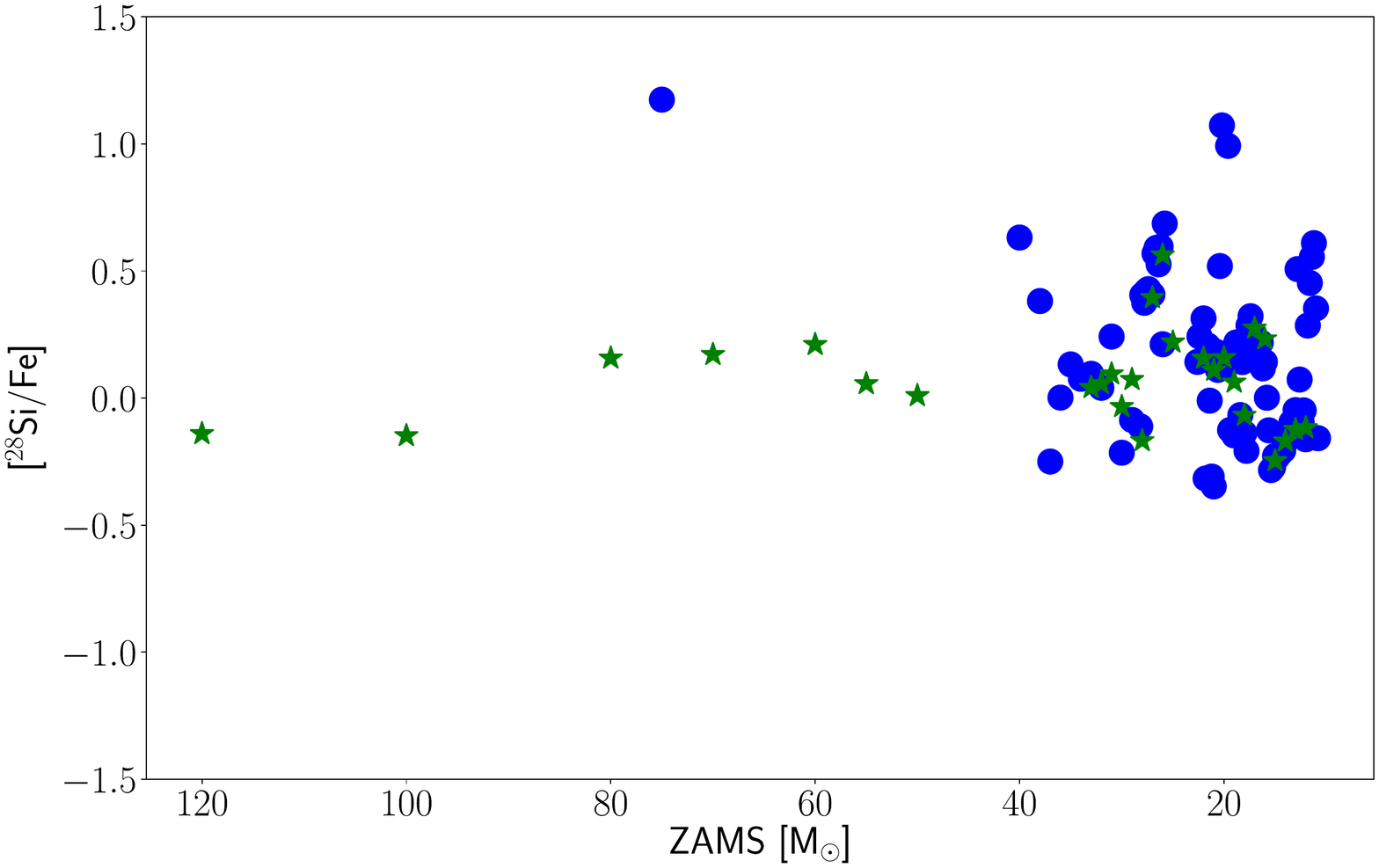} \\
\includegraphics[width=0.49\textwidth]{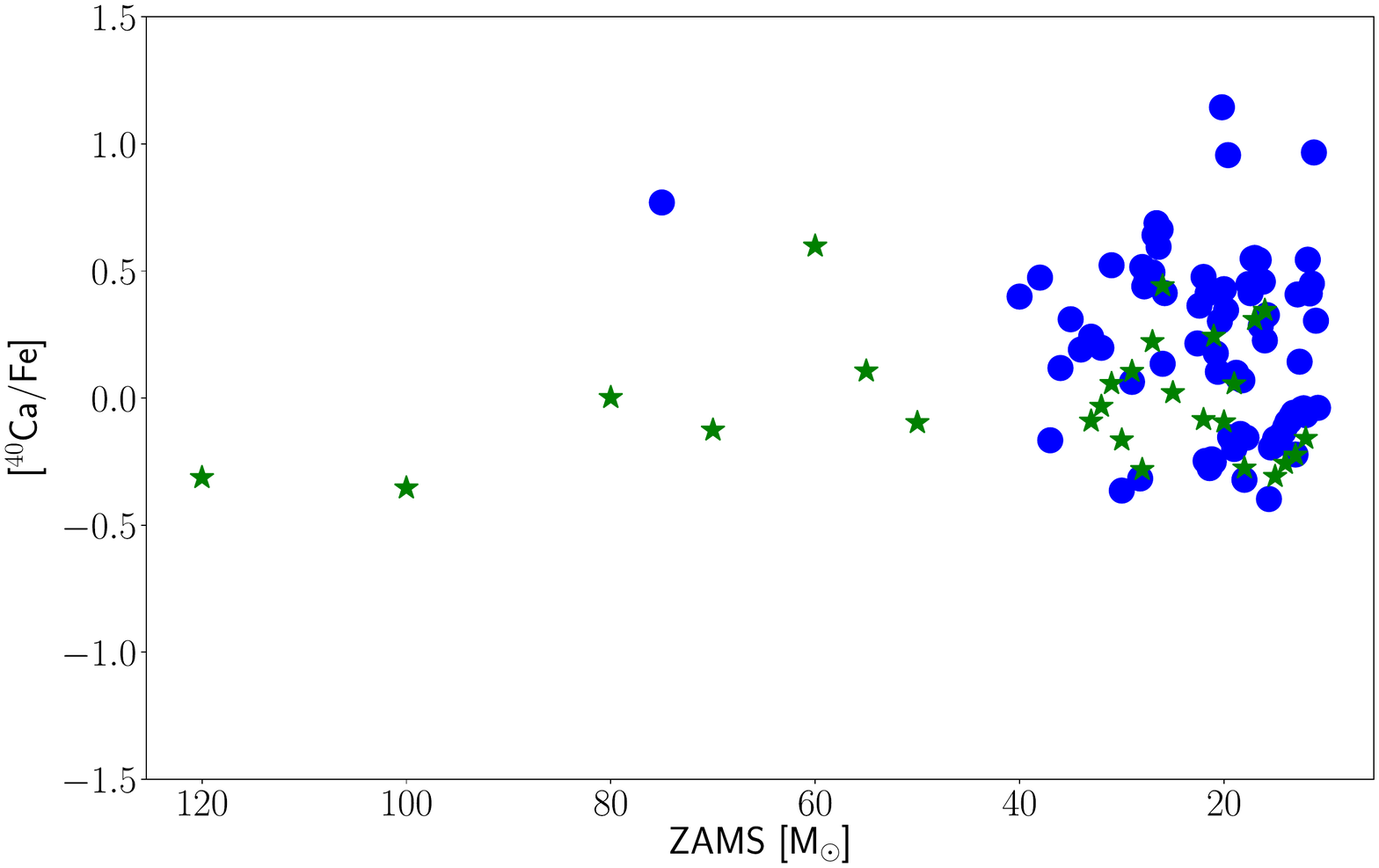}	
\end{tabular}
\caption{Top to bottom: Abundance ratios of $^{16}$O, $^{28}$Si and $^{40}$Ca to Fe (relative to the solar ratio) for the WHW02 models (blue circles) and WH07 models (green stars). 
	\label{fig:ratios_alphas}
    } 
\end{center}
\end{figure}

%%%%%%%%%%%%%%%%%%%%%%%%%%%%%%%%%%%%%%%%%%%%%%%%%%%
%% Section 5: Summary
%%%%%%%%%%%%%%%%%%%%%%%%%%%%%%%%%%%%%%%%%%%%%%%%%%%
\section{Summary and Discussion} 
\label{sec:summary}

Core-collapse supernovae represent a fascinating problem that still suffers large uncertainties despite decades of work. The detailed explosion mechanism is an important open question and its solution requires computationally expensive simulations. The unsolved explosion mechanism problem and the high computational cost of multi-dimensional simulations pose a problem for nucleosynthesis predictions. This warrants effective approaches to predict CCSN nucleosynthesis yields that go beyond the traditional piston or thermal bomb methods with their well-known limitations.

In this work, we used the successful explosions from the standard PUSH calibration of \cite{push2} to compute the nucleosynthesis yields from CCSNe. These explosions were obtained in spherical symmetry using the PUSH method, which was first introduced in \cite{push1} and calibrated in \cite{push2}. 
The calibration of PUSH was compared to results of multi-dimensional simulations of \cite{pan16,bruenn16,ebinger.phd}, in order to ensure a satisfactory agreement with models that are able to capture the possible supernova mechanism self-consistently. Since PUSH is a parametrized method, we cannot capture fundamentally multidimensional behavior like simultaneous downflow and outflow of matter which can result in different entropies for regions that are at similar distances from the PNS. We also cannot calculate the fallback since early time fallback can be determined only in multidimensional simulations whereas late fallback requires simulation times much longer than what is feasible with our setup. However, with respect to the temporal evolution of the shock radius, neutrino heating rates and entropy profiles, our setup enables explosions that are in good agreement with outcomes of multi-dimensional simulations.

We post-process all 111 models that successfully exploded with a nuclear reaction network to obtain detailed nucleosynthesis yields. Except for one model (the calibration model which was required to reproduce SN~1987A), the yields (including the iron-group yields), the explosion energy, and the location of the mass cut are self-consistent predictions (see Paper~II for a detailed discussion of this).

The main results and findings from this study are:
\begin{itemize}
\item Our nucleosynthesis yields include predictions for iron-group elements based on a mass cut consistent with the explosion energy and a consistent treatment of neutrino interactions with matter. We present a detailed discussion of the synthesis mechanisms of iron-group and intermediate mass elements.  
\item In our models, the $Y_e$ in the innermost layers is different from the pre-explosion value. In terms of mass coordinate (moving outward from the mass cut), we encounter first the late neutrino wind ($Y_e \sim$0.42), then proton-rich layers ($Y_e >$0.5), followed by layers with slightly lower $Y_e$ compared to its pre-explosion value. Further out, the $Y_e$ remains unaffected by the explosion. Synthesis of isotopes beyond iron occurs in the neutrino wind as well as in the proton-rich layers, although the former dominates their yields.
\item Multi-dimensional simulations show a variety of $Y_e$ values: There exist outer layers shocked by a close-to-spherical shock wave, where $Y_e$ is either the original one from the progenitor model or only slightly changed (lowered) by beta-decays of explosive burning products and electron captures during explosive burning; there exist layers which experience strong neutrino and anti neutrino absorption, causing $Y_e>0.5$ and leading to a $\nu$p-process (see e.g.\ \citet{Eichler.Nakamura.ea:2017}); and convective blobs from deeper inside can convect neutron-rich layers outwards,  
leading to weakly neutron-rich conditions (see e.g.\ \citet{Harris.Hix.ea:2017}). With our PUSH approach, we obtain a similar range of $Y_e$ values as found in full multi-dimensional simulations. 
The amount of $\nu$p-process and weak r-process elements are the first predictions from spherically symmetric simulations which take neutrino transport consistently into account. The comparison of our results to observations of elemental and isotopic composition of iron-group elements (see Section~\ref{sec:observations}) puts constraints on the question whether the relative ratios are consistent and whether the amount of $\nu$p-process and weak r-process abundances relate to reality.
\item We compare in detail two representative models, s16.0 and s21.0. We find that the location of the mass cut is related to the compactness and impacts the relative amount of matter undergoing complete Si-burning versus incomplete Si-burning. This, in addition to the explosion energy, influences the total yields.
\item We find several interesting trends of isotopic and elemental yields with compactness. Symmetric iron-group elements (such as $^{56}$Ni or $^{44}$Ti) exhibit linear trends, while asymmetric isotopes such as $^{57}$Ni or $^{58}$Ni depend more strongly on the electron fraction $Y_e$. Similar relationships are also seen in elemental yields of iron-group elements, mostly determined by the isotope with the largest contribution to the elemental yield. $^{56}$Ni (decaying to $^{56}$Fe) and $^{44}$Ti are nuclei dependent on the strength of the explosion as well as on the mass cut between the neutron star remnant and ejecta. Asymmetric (more neutron-rich) Ni/Fe-isotopes measure also the neutron excess or $Y_e$ of the innermost (Fe-group) ejecta.
\item Our predicted yields of $^{56}$Ni from the PUSH method are in good agreement with observationally derived values for CCSNe. 
\item We compare our results for the iron-group nuclei to observationally derived abundances of a metal-poor star. Overall, we find good agreement. However, for some elements (Sc and Zn, for example) the models show large variations. When comparing the yields from the two sets of models, we find that the details of the pre-explosion compositional structure impacts the explosive yields, in particular for odd-$Z$ elements in the iron group (for example, Mn).
\item We compare our results for alpha elements with observational [$\alpha$/Fe] trends. Our overall results are in good agreement for [Ca/Fe] and [Si/Fe], whereas only stars above 20~M$_\odot$ reproduce the typical [O/Fe] ratios. However, all our models have solar metallicity and a true comparison requires yields for low-metallicity models as well. Such models will be explored in a future study.
\item The real test of our models is to investigate the impact of our yields on chemical evolution simulations. Machine readable table of isotopic yields for all exploding models of the WH07 set are included with this Paper. Isotopic yields for the exploding models of the WHW02 set are available upon request.
\end{itemize}

\acknowledgments
The work at NC State (SC, KE, CF) was supported by the Department of Energy through an Early CAREER Award (DOE grant No.SC0010263).
CF acknowledges support from the Research Corporation for Science Advancement through a Cottrell Scholar Award.
The effort at the Universit\"at Basel was supported by the Schweizerischer Nationalfond and by the ERC Advanced Grant ``FISH''.
KE acknowledges support from the GSI.
AP acknowledges support from the INFN project "High Performance data Network" funded by the Italian CIPE.

\pagebreak
\bibliographystyle{yahapj}
\bibliography{references_push,references_fkt}

\appendix

\section{Table of key isotopes} \label{appx:key_iso_all}
Table~\ref{tab:nitisample} gives the isotopic yields of $^{56-58}$Ni and $^{44}$Ti for all exploding models of this paper.

\begin{table*}
	\begin{center}
		\caption{Isotopic yields in M~$_{\odot}$ for all models
        	\label{tab:nitisample}
        }
		\begin{tabular}{llllllllll}
			\tableline \tableline
			Model & $^{56}$Ni & $^{57}$Ni & $^{58}$Ni & $^{44}$Ti & Model & $^{56}$Ni & $^{57}$Ni & $^{58}$Ni & $^{44}$Ti\\
			(-) & (M$_{\odot}$)  & (M$_{\odot}$)  & (M$_{\odot}$) & (M$_{\odot}$) & (-) & (M$_{\odot}$)  & (M$_{\odot}$)  & (M$_{\odot}$) & (M$_{\odot}$) \\
			\tableline
		s10.8  &  2.71E-02  &  5.25E-04  &  1.82E-04  & 1.45E-05 & w12.0   & 3.87E-02  & 1.33E-03  & 1.38E-03 &  2.37E-05\\
		s11.0  &  2.38E-02  &  3.46E-04  &  1.25E-04  &  1.15E-05 & w13.0   & 5.50E-02  & 1.92E-03  & 2.09E-03  & 3.48E-05\\
		s11.2  &  2.31E-02  &  3.12E-04  &  1.01E-04  &  1.28E-05 & w14.0   & 7.15E-02 &  2.51E-03 &  2.85E-03 &    4.26E-05\\
		s11.4  &  2.24E-02  &  3.13E-04  &  1.08E-04  &  1.08E-05 & w15.0  & 1.17E-01  & 4.28E-03 &  7.60E-03   & 5.38E-05\\
		s11.6  &  2.43E-02  &  3.34E-04  &  1.14E-04  &  1.08E-05 & w16.0  & 6.64E-02  & 2.09E-03  & 2.88E-03   & 3.02E-05\\
		s11.8  &  2.39E-02  &  3.70E-04  &  1.27E-04  &  1.36E-05 & w17.0  & 7.03E-02  & 2.18E-03 &  2.87E-03  & 3.09E-05\\
		s12.0  &  3.69E-02  &  1.51E-03  &  3.58E-03  &  2.06E-05 & w18.0  & 9.40E-02  & 3.58E-03  & 6.32E-03   & 5.87E-05\\
		s12.2  &  3.38E-02  &  7.28E-04  &  3.77E-04  &  2.32E-05 & w19.0  & 8.71E-02  & 2.69E-03  & 3.26E-03   & 3.57E-05\\
		s12.4  &  4.06E-02  &  6.86E-04  &  2.46E-04  &  2.33E-05 & w20.0  & 1.12E-01  & 3.41E-03  & 4.24E-03   & 4.92E-05\\
        s12.6  &  4.76E-02  &  7.00E-04  &  3.56E-04  &  2.20E-05 & w21.0  & 6.22E-02  & 2.46E-03  & 6.07E-03   & 2.98E-05\\

			\tableline
		\end{tabular}
	\end{center}
    \tablecomments{Table \ref{tab:nitisample} is published in its entirety in the machine-readable format. A portion is shown here for guidance regarding its form and content.}
	%\tablecomments{The table columns are: zero age main sequence \textbf{(ZAMS)} mass, compactness at bounce, total mass at collapse, mass of the iron core, carbon-oxygen core, and helium core, mass of the hydrogen-rich envelope, explosion energy ($1\; {\rm B}= 1\;{\rm Bethe} = 10^{51}~\rm erg$) and remnant mass. The column after the remnant mass indicates which layer in the pre-explosion structure this corresponds to. The last four columns give the nucleosynthesis yields of observable isotopes of nickel and titanium. 
%	}
\end{table*}

\section{Table of long-lived radionuclides}
\label{appx:radionuclides}
Table~\ref{tab:radiosample} gives the total yields of key radionuclides before decay to stability for all exploding models of this paper. The yields provided include pre-explosion and explosive contributions.

\begin{table*}
	\begin{center}
		\caption{Total pre-decay yields of selected long-lived radionuclides in M~$_{\odot}$ for all models
        	\label{tab:radiosample}
        }
		\begin{tabular}{lllllllllll}
			\tableline \tableline
			Model & $^{26}$Al & $^{41}$Ca & $^{44}$Ti & $^{48}$V & $^{53}$Mn & $^{60}$Fe & $^{81}$Kr & $^{93}$Zr & $^{97}$Tc & $^{98}$Tc\\
			(-) & (M$_{\odot}$)  & (M$_{\odot}$)  & (M$_{\odot}$) & (M$_{\odot}$) & (M$_{\odot}$)  & (M$_{\odot}$)  & (M$_{\odot}$) & (M$_{\odot}$) & (M$_{\odot}$) & (M$_{\odot}$)\\
			\tableline
s10.8 & 3.28E-07 & 2.37E-07 & 1.45E-05 & 1.98E-08 & 1.06E-06 & 2.97E-08 & 6.21E-09 & 2.36E-09 & 1.84E-13 & 6.11E-16\\
s11.0 & 2.35E-07 & 8.79E-07 & 1.15E-05 & 2.15E-09 & 1.73E-06 & 6.40E-08 & 5.77E-11 & 1.19E-08 & 3.74E-15 & 9.25E-17\\
s11.2 & 1.85E-07 & 8.70E-07 & 1.28E-05 & 1.19E-08 & 1.46E-06 & 8.52E-08 & 2.12E-08 & 5.78E-08 & 1.73E-13 & 4.73E-16\\
s11.4 & 2.00E-07 & 6.69E-07 & 1.08E-05 & 2.34E-09 & 1.52E-06 & 6.57E-08 & 2.91E-09 & 1.84E-08 & 6.18E-14 & 2.52E-16\\
s11.6 & 2.06E-07 & 8.12E-07 & 1.08E-05 & 5.04E-09 & 1.58E-06 & 1.79E-07 & 7.57E-09 & 1.93E-08 & 1.05E-13 & 2.11E-15\\
s11.8 & 2.16E-07 & 6.03E-07 & 1.36E-05 & 4.61E-08 & 1.37E-06 & 2.96E-08 & 4.86E-08 & 1.24E-08 & 2.50E-13 & 7.22E-16\\
s12.0 & 7.34E-08 & 3.25E-07 & 2.06E-05 & 4.43E-09 & 4.31E-06 & 2.42E-08 & 3.67E-20 & 6.32E-09 & 1.23E-22 & 6.90E-20\\
s12.2 & 2.89E-07 & 3.09E-07 & 2.32E-05 & 8.50E-08 & 2.19E-06 & 2.87E-08 & 6.42E-08 & 6.75E-09 & 3.12E-13 & 1.71E-15\\
s12.4 & 3.92E-07 & 4.59E-07 & 2.33E-05 & 3.89E-09 & 2.64E-06 & 4.68E-08 & 2.83E-19 & 1.31E-08 & 6.47E-19 & 1.77E-17\\
s12.6 & 2.89E-07 & 1.15E-06 & 2.20E-05 & 3.04E-09 & 2.80E-06 & 4.05E-08 & 1.97E-11 & 2.37E-08 & 1.76E-15 & 5.66E-17\\
%s12.8 & 2.21E-07 & 7.91E-07 & 1.36E-05 & 3.18E-08 & 2.32E-06 & 6.83E-08 & 4.13E-08 & 2.92E-08 & 1.71E-13 & 5.77E-16\\
%s13.0 & 7.76E-08 & 2.34E-06 & 2.07E-05 & 3.19E-08 & 2.01E-05 & 3.67E-08 & 4.67E-09 & 1.98E-08 & 9.36E-14 & 2.58E-16\\
%& & & & \\
%w12.0 & 1.04E-05 & 1.63E-06 & 2.37E-05 & 1.34E-08 & 7.65E-06 & 6.44E-06 & 3.34E-09 & 7.50E-09 & 3.08E-11 & 1.32E-11\\
%w13.0 & 1.84E-05 & 2.20E-06 & 3.48E-05 & 1.34E-08 & 1.09E-05 & 1.83E-05 & 1.97E-09 & 1.49E-08 & 3.04E-11 & 1.50E-11\\
%w14.0 & 3.13E-05 & 2.63E-06 & 4.26E-05 & 1.68E-08 & 1.39E-05 & 2.81E-05 & 2.29E-09 & 5.04E-08 & 3.27E-11 & 1.60E-11\\
%w15.0 & 2.18E-05 & 3.49E-06 & 5.38E-05 & 3.75E-08 & 1.87E-05 & 2.32E-05 & 1.18E-08 & 1.39E-08 & 3.72E-11 & 1.72E-11\\
%w16.0 & 1.55E-05 & 1.97E-05 & 3.02E-05 & 1.60E-07 & 1.95E-05 & 4.09E-05 & 5.18E-08 & 2.05E-08 & 3.94E-11 & 2.04E-11\\
%w17.0 & 1.54E-05 & 2.03E-05 & 3.09E-05 & 1.11E-07 & 2.05E-05 & 5.30E-05 & 7.67E-09 & 2.88E-08 & 4.04E-11 & 2.13E-11\\
			\tableline
		\end{tabular}
	\end{center}
    \tablecomments{Table \ref{tab:radiosample} is published in its entirety in the machine-readable format. A portion is shown here for guidance regarding its form and content.}
\end{table*}

\section{Tables of complete isotopic yields} \label{appx:tables}
Table~\ref{tab:finabs02} gives the detailed isotopic composition of the post-processed ejecta
for three models from the WHW02 set. Models s16.0 and s21.0 are discussed in detail in the paper while model s18.8 is the best fit to SN1987A.
Only material that reaches peak temperatures $\sim 1.75$~GK is post-processed with the nuclear reaction network and listed in this Table. This corresponds to 0.73~\msun (s16.0), 0.79~\msun (s18.8), and 0.91~\msun (s21.0) of material above the mass cut. 
The contributions from outer (unaltered) layers of the pre-explosion model to the total ejecta are not included, but they are given in the machine readable tables.

\begin{center}
\begin{longtable*}{|c|lll|c|lll|}

\multicolumn{8}{c}%
{{\bfseries \tablename\ \thetable{} -- Isotopic yields in M~$_{\odot}$ for selected WHW02 models}} \label{tab:finabs02}\\

\hline
\multicolumn{1}{|c|}{\textbf{Model}} & \multicolumn{1}{c}{\textbf{s16.0}} & \multicolumn{1}{c}{\textbf{s18.8}} & \multicolumn{1}{c|}{\textbf{s21.0}} & \multicolumn{1}{|c|}{\textbf{Model}} & \multicolumn{1}{c}{\textbf{s16.0}} & \multicolumn{1}{c}{\textbf{s18.8}} & \multicolumn{1}{c|}{\textbf{s21.0}} \\ 

\endfirsthead

\multicolumn{8}{c}%
{{\bfseries \tablename\ \thetable{} -- continued from previous page}} \\
\hline
\multicolumn{1}{|c|}{\textbf{Model}} & \multicolumn{1}{c}{\textbf{s16.0}} & \multicolumn{1}{c}{\textbf{s18.8}} & \multicolumn{1}{c|}{\textbf{s21.0}} & \multicolumn{1}{|c|}{\textbf{Model}} & \multicolumn{1}{c}{\textbf{s16.0}} & \multicolumn{1}{c}{\textbf{s18.8}} & \multicolumn{1}{c|}{\textbf{s21.0}} \\ 
\hline 
\endhead
\hline

\endfoot

\hline
\endlastfoot 

 \tableline
$^{1}$H & 1.83E-04 & 1.81E-04 & 2.33E-04 & $^{37}$Cl & 3.82E-04 & 4.11E-04 & 1.74E-06 \\
$^{2}$H & 5.03E-18 & 5.09E-18 & 1.25E-18 & $^{36}$Ar & 1.82E-02 & 1.57E-02 & 5.35E-03 \\
$^{3}$He & 2.30E-12 & 1.54E-12 & 2.57E-12 & $^{38}$Ar & 3.26E-03 & 3.13E-03 & 1.16E-06 \\
$^{4}$He & 1.58E-02 & 1.52E-02 & 2.58E-02 & $^{40}$Ar & 4.29E-06 & 5.54E-06 & 2.68E-08 \\
$^{6}$Li & 1.10E-19 & 1.45E-19 & 2.25E-20 & $^{39}$K & 2.21E-04 & 1.84E-04 & 5.48E-06 \\
$^{7}$Li & 4.83E-11 & 4.02E-11 & 6.69E-11 & $^{41}$K & 5.39E-05 & 3.71E-05 & 7.41E-07 \\
$^{9}$Be & 6.60E-21 & 6.87E-21 & 4.47E-21 & $^{40}$Ca & 1.02E-02 & 6.61E-03 & 5.11E-03 \\
$^{10}$B & 1.87E-16 & 1.55E-16 & 4.31E-16 & $^{42}$Ca & 1.15E-04 & 9.43E-05 & 4.33E-07 \\
$^{11}$B & 1.93E-11 & 1.42E-11 & 3.09E-11 & $^{43}$Ca & 8.58E-07 & 8.13E-07 & 7.21E-07 \\
$^{12}$C & 3.49E-03 & 1.21E-03 & 5.48E-03 & $^{44}$Ca & 3.27E-05 & 3.06E-05 & 6.18E-05 \\
$^{13}$C & 3.85E-10 & 3.63E-10 & 6.25E-10 & $^{46}$Ca & 1.56E-07 & 1.86E-07 & 1.14E-07 \\
$^{14}$N & 1.85E-08 & 8.02E-09 & 2.42E-08 & $^{45}$Sc & 1.22E-06 & 9.94E-07 & 1.30E-06 \\
$^{15}$N & 4.20E-08 & 2.60E-08 & 7.91E-08 & $^{46}$Ti & 4.19E-05 & 3.13E-05 & 6.84E-07 \\
$^{16}$O & 2.51E-01 & 3.08E-01 & 4.81E-01 & $^{47}$Ti & 3.23E-06 & 3.04E-06 & 2.56E-06 \\
$^{17}$O & 6.46E-09 & 6.20E-09 & 8.30E-09 & $^{48}$Ti & 9.85E-05 & 9.15E-05 & 1.61E-04 \\
$^{18}$O & 8.39E-10 & 6.11E-10 & 7.34E-10 & $^{49}$Ti & 5.07E-06 & 4.56E-06 & 3.04E-06 \\
$^{19}$F & 3.85E-11 & 3.76E-11 & 5.15E-11 & $^{50}$Ti & 6.96E-07 & 7.66E-07 & 2.94E-07 \\
$^{20}$Ne & 7.53E-02 & 8.37E-02 & 9.71E-02 & $^{51}$V & 1.79E-05 & 1.60E-05 & 1.27E-05 \\
$^{21}$Ne & 2.76E-07 & 3.38E-08 & 3.60E-07 & $^{50}$Cr & 1.52E-04 & 1.37E-04 & 2.01E-06 \\
$^{22}$Ne & 1.26E-06 & 7.93E-07 & 2.09E-06 & $^{52}$Cr & 7.70E-04 & 7.24E-04 & 1.23E-03 \\
$^{23}$Na & 1.05E-05 & 2.36E-06 & 1.39E-05 & $^{53}$Cr & 1.13E-04 & 1.06E-04 & 3.17E-05 \\
$^{24}$Mg & 2.35E-02 & 3.13E-02 & 4.48E-02 & $^{54}$Cr & 5.14E-07 & 6.47E-07 & 1.39E-07 \\
$^{25}$Mg & 1.81E-06 & 2.09E-06 & 1.36E-06 & $^{55}$Mn & 6.37E-04 & 6.02E-04 & 1.01E-04 \\
$^{26}$Mg & 3.07E-06 & 4.34E-06 & 3.67E-07 & $^{54}$Fe & 8.42E-03 & 7.68E-03 & 6.85E-04 \\
$^{27}$Al & 1.34E-04 & 1.55E-04 & 1.02E-04 & $^{56}$Fe & 7.58E-02 & 6.94E-02 & 1.36E-01 \\
$^{28}$Si & 1.33E-01 & 1.42E-01 & 6.35E-02 & $^{57}$Fe & 2.98E-03 & 2.67E-03 & 2.85E-03 \\
$^{29}$Si & 3.82E-04 & 4.96E-04 & 3.74E-05 & $^{58}$Fe & 4.89E-07 & 6.25E-07 & 1.49E-07 \\
$^{30}$Si & 4.58E-04 & 6.24E-04 & 9.43E-06 & $^{59}$Co & 1.52E-04 & 1.36E-04 & 3.93E-04 \\
$^{31}$P & 5.24E-04 & 6.88E-04 & 1.55E-05 & $^{58}$Ni & 7.41E-03 & 6.56E-03 & 1.85E-03 \\
$^{32}$S & 7.42E-02 & 6.70E-02 & 2.59E-02 & $^{60}$Ni & 1.50E-03 & 1.42E-03 & 3.37E-03 \\
$^{33}$S & 5.14E-04 & 5.18E-04 & 1.05E-05 & $^{61}$Ni & 8.17E-05 & 7.48E-05 & 1.07E-04 \\
$^{34}$S & 4.16E-03 & 4.64E-03 & 6.20E-06 & $^{62}$Ni & 9.66E-04 & 1.03E-03 & 3.56E-04 \\
$^{36}$S & 2.01E-03 & 2.24E-03 & 3.64E-08 & $^{64}$Ni & 2.61E-06 & 1.71E-06 & 5.78E-07 \\
$^{35}$Cl & 2.56E-04 & 2.84E-04 & 1.06E-05 & $^{63}$Cu & 2.84E-06 & 6.49E-06 & 2.46E-05 \\
& & & & & & & \\
& & & & & & & \\
$^{65}$Cu & 1.24E-06 & 3.72E-06 & 4.52E-06 & $^{93}$Nb & 3.05E-04 & 3.63E-04 & 2.20E-04  \\                               
$^{64}$Zn & 9.59E-06 & 1.44E-05 & 2.36E-04 & $^{92}$Mo & 2.02E-17 & 1.01E-08 & 1.07E-07  \\                               
$^{66}$Zn & 1.61E-05 & 1.73E-04 & 1.84E-04 & $^{94}$Mo & 2.50E-12 & 5.71E-09 & 9.75E-09  \\                               
$^{67}$Zn & 1.78E-06 & 2.53E-06 & 2.54E-06 & $^{95}$Mo & 9.08E-06 & 1.34E-05 & 3.74E-06  \\                               
$^{68}$Zn & 3.73E-06 & 1.00E-05 & 3.41E-06 & $^{96}$Mo & 2.71E-11 & 1.71E-11 & 6.63E-12  \\                                
$^{70}$Zn & 3.65E-06 & 2.21E-06 & 7.69E-07 & $^{97}$Mo & 4.09E-05 & 5.13E-05 & 3.14E-05  \\                                
$^{69}$Ga & 3.09E-06 & 4.61E-06 & 2.62E-06 & $^{98}$Mo & 1.43E-05 & 1.66E-05 & 1.29E-05  \\
$^{71}$Ga & 1.95E-06 & 2.18E-06 & 1.41E-06 & $^{100}$Mo & 2.36E-05 & 3.19E-05 & 8.52E-06  \\
$^{70}$Ge & 9.44E-08 & 1.52E-05 & 3.55E-05 & $^{96}$Ru & 4.14E-19 & 7.85E-14 & 1.97E-12 \\
$^{72}$Ge & 3.91E-06 & 2.46E-05 & 9.99E-06 & $^{98}$Ru & 2.99E-17 & 4.05E-13 & 1.41E-12 \\
$^{73}$Ge & 5.82E-07 & 8.07E-07 & 6.85E-07 & $^{99}$Ru & 5.80E-06 & 6.53E-06 & 3.25E-06 \\
$^{74}$Ge & 3.18E-06 & 2.75E-06 & 8.89E-07 & $^{100}$Ru & 1.45E-13 & 1.79E-13 & 1.77E-13  \\
$^{76}$Ge & 9.19E-06 & 6.08E-06 & 2.00E-06 & $^{101}$Ru & 2.25E-05 & 2.60E-05 & 1.26E-05  \\
$^{75}$As & 2.76E-06 & 3.60E-06 & 1.13E-06 & $^{102}$Ru & 1.48E-04 & 2.12E-04 & 5.31E-05  \\
$^{74}$Se & 4.48E-10 & 1.32E-07 & 1.22E-06 & $^{104}$Ru & 9.18E-04 & 1.00E-03 & 3.58E-04  \\
$^{76}$Se & 1.75E-10 & 5.63E-06 & 7.09E-06 & $^{103}$Rh & 8.57E-05 & 1.11E-04 & 4.03E-05  \\
$^{77}$Se & 1.02E-05 & 1.03E-05 & 5.94E-06 & $^{102}$Pd & 2.16E-24 & 1.25E-17 & 9.49E-17  \\
$^{78}$Se & 2.31E-06 & 6.27E-06 & 2.06E-06 & $^{104}$Pd & 6.05E-15 & 8.42E-15 & 5.64E-15  \\
$^{80}$Se & 1.74E-05 & 1.60E-05 & 9.70E-06 & $^{105}$Pd & 6.78E-04 & 8.27E-04 & 2.30E-04  \\
$^{82}$Se & 4.32E-05 & 4.22E-05 & 2.07E-05 & $^{106}$Pd & 2.31E-04 & 3.11E-04 & 7.07E-05  \\
$^{79}$Br & 1.23E-05 & 1.42E-05 & 8.57E-06 & $^{108}$Pd & 2.34E-04 & 2.47E-04 & 8.76E-05  \\
$^{81}$Br & 3.05E-05 & 3.17E-05 & 1.78E-05 & $^{110}$Pd & 1.98E-04 & 1.94E-04 & 6.32E-05  \\
$^{78}$Kr & 3.85E-12 & 1.72E-09 & 4.37E-08 & $^{107}$Ag & 2.67E-04 & 2.60E-04 & 8.28E-05  \\
$^{80}$Kr & 2.04E-12 & 8.10E-08 & 5.75E-07 & $^{109}$Ag & 1.68E-04 & 1.88E-04 & 6.15E-05  \\
$^{82}$Kr & 4.45E-10 & 8.31E-07 & 9.84E-07 & $^{106}$Cd & 0.00E+00 & 6.55E-23 & 9.88E-22  \\
$^{83}$Kr & 1.96E-05 & 1.69E-05 & 1.09E-05 & $^{108}$Cd & 2.53E-22 & 8.59E-22 & 8.02E-22  \\
$^{84}$Kr & 1.21E-04 & 6.18E-05 & 1.41E-05 & $^{110}$Cd & 3.77E-13 & 5.35E-13 & 2.19E-13  \\
$^{86}$Kr & 2.29E-04 & 8.00E-04 & 4.98E-05 & $^{111}$Cd & 2.16E-04 & 2.34E-04 & 8.59E-05  \\
$^{85}$Rb & 6.36E-06 & 1.55E-05 & 5.60E-06 & $^{112}$Cd & 1.91E-04 & 2.36E-04 & 7.85E-05  \\
$^{87}$Rb & 9.37E-05 & 7.18E-05 & 2.63E-06 & $^{113}$Cd & 7.88E-05 & 9.45E-05 & 2.62E-05  \\
$^{84}$Sr & 2.07E-15 & 2.33E-10 & 1.31E-08 & $^{114}$Cd & 1.96E-04 & 2.27E-04 & 6.10E-05  \\
$^{86}$Sr & 1.73E-09 & 2.86E-07 & 2.06E-07 & $^{116}$Cd & 9.96E-05 & 1.24E-04 & 3.27E-05  \\
$^{87}$Sr & 4.09E-10 & 1.97E-07 & 6.83E-08 & $^{113}$In & 1.66E-22 & 5.18E-22 & 1.60E-22  \\
$^{88}$Sr & 7.41E-04 & 2.97E-04 & 1.54E-05 & $^{115}$In & 1.08E-05 & 1.33E-05 & 4.63E-06  \\
$^{89}$Y & 4.74E-05 & 1.04E-04 & 3.02E-05  & $^{114}$Sn & 4.83E-20 & 1.36E-19 & 5.48E-20  \\
$^{90}$Zr & 8.25E-04 & 9.02E-04 & 5.25E-04 & $^{115}$Sn & 3.55E-23 & 9.54E-23 & 4.38E-23  \\
$^{91}$Zr & 4.72E-05 & 5.01E-05 & 1.50E-05 & $^{116}$Sn & 9.81E-16 & 1.36E-15 & 6.85E-16  \\
$^{92}$Zr & 2.66E-05 & 2.74E-05 & 1.79E-05 & $^{117}$Sn & 9.37E-05 & 1.15E-04 & 4.35E-05  \\
$^{94}$Zr & 5.92E-04 & 6.57E-04 & 2.79E-04 & $^{118}$Sn & 3.13E-05 & 3.90E-05 & 1.80E-05  \\
$^{96}$Zr & 5.85E-05 & 4.85E-05 & 5.14E-05 & $^{119}$Sn & 4.99E-06 & 6.03E-06 & 4.39E-06  \\
$^{120}$Sn & 1.76E-05 & 2.34E-05 & 2.28E-05 & $^{135}$Ba & 3.03E-13 & 5.02E-13 & 5.64E-09\\
$^{122}$Sn & 7.25E-06 & 8.39E-06 & 9.97E-06 & $^{136}$Ba & 3.20E-23 & 7.54E-23 & 5.95E-19\\
$^{124}$Sn & 6.67E-06 & 1.05E-05 & 6.70E-06 & $^{137}$Ba & 2.89E-16 & 8.06E-16 & 4.60E-08\\
$^{121}$Sb & 1.02E-05 & 1.45E-05 & 1.72E-05 & $^{138}$Ba & 6.89E-15 & 1.36E-14 & 3.31E-07 \\
$^{123}$Sb & 8.58E-06 & 1.04E-05 & 1.20E-05 & $^{139}$La & 1.57E-15 & 3.97E-15 & 6.20E-08 \\
$^{120}$Te & 6.91E-25 & 2.09E-24 & 1.91E-24 & $^{140}$Ce & 5.10E-17 & 8.22E-17 & 2.30E-08 \\
$^{122}$Te & 9.09E-14 & 1.25E-13 & 3.35E-13 & $^{142}$Ce & 1.11E-16 & 3.46E-16 & 2.37E-08\\
$^{123}$Te & 1.74E-21 & 2.34E-21 & 9.99E-21 & $^{141}$Pr & 6.11E-16 & 8.16E-16 & 1.62E-08\\
$^{124}$Te & 3.88E-15 & 4.80E-15 & 8.79E-15 & $^{142}$Nd & 0.00E+00 & 0.00E+00 & 1.69E-22\\
$^{125}$Te & 2.37E-05 & 3.32E-05 & 2.35E-05 & $^{143}$Nd & 6.36E-18 & 1.48E-17 & 1.16E-08\\
$^{126}$Te & 5.67E-05 & 7.36E-05 & 7.19E-05 & $^{144}$Nd & 2.96E-16 & 5.92E-16 & 1.61E-09\\
$^{128}$Te & 1.99E-05 & 2.66E-05 & 4.00E-05 & $^{145}$Nd & 1.80E-17 & 5.55E-17 & 3.18E-08\\
$^{130}$Te & 2.77E-04 & 3.17E-04 & 1.89E-04 & $^{146}$Nd & 3.52E-18 & 9.29E-18 & 3.36E-09\\
$^{127}$I & 1.10E-05 & 1.47E-05 & 2.62E-05 & $^{148}$Nd & 6.60E-18 & 1.18E-17 & 3.97E-08\\
$^{128}$Xe & 6.59E-20 & 9.06E-20 & 3.65E-19 & $^{150}$Nd & 1.56E-17 & 4.16E-17 & 1.38E-08\\
$^{129}$Xe & 1.12E-04 & 1.51E-04 & 1.36E-04 & $^{147}$Sm & 5.94E-17 & 1.39E-16 & 7.26E-08\\
$^{130}$Xe & 2.67E-15 & 3.46E-15 & 3.44E-15 & $^{148}$Sm & 0.00E+00 & 0.00E+00 & 9.12E-18\\
$^{131}$Xe & 3.03E-04 & 3.40E-04 & 1.46E-04 & $^{149}$Sm & 1.44E-17 & 1.72E-17 & 3.72E-08\\
$^{132}$Xe & 2.08E-04 & 2.35E-04 & 9.78E-05 & $^{150}$Sm & 0.00E+00 & 0.00E+00 & 1.88E-14\\
$^{134}$Xe & 8.73E-10 & 1.05E-09 & 7.03E-06 & $^{152}$Sm & 7.23E-18 & 1.11E-17 & 7.27E-09\\
$^{136}$Xe & 4.79E-15 & 1.14E-14 & 7.71E-08 & $^{154}$Sm & 7.41E-19 & 1.85E-18 & 9.99E-09\\
$^{133}$Cs & 1.04E-06 & 1.23E-06 & 1.93E-05 & $^{151}$Eu & 1.31E-18 & 3.31E-18 & 1.91E-08\\
$^{134}$Ba & 8.24E-22 & 1.07E-21 & 2.01E-19 & $^{153}$Eu & 4.48E-18 & 1.04E-17 & 1.41E-08\\

\end{longtable*}
\end{center}

Table~\ref{tab:finabw07} gives the detailed isotopic composition of the post-processed ejecta and of the additional unprocessed material from the pre-explosion model, for all models of the WH07 sample\footnote{http://astro.physics.ncsu.edu/$\sim$cfrohli/}. Explosive yields for the WHW02 sample can be provided upon request.

\begin{table*}
	\begin{center}
		\caption{Isotopic yields in M~$_{\odot}$ for the WH07 set
        	\label{tab:finabw07}
        }
		\begin{tabular}{llll}
			\tableline \tableline
			Model & Isotope & Explosive yields & Progenitor yields \\
			(-) & (-)   & (M$_{\odot}$)  & (M$_{\odot}$)  \\
			\tableline
		w12.0	& $^{1}$H  & 8.98E-05 & 5.27E+00\\
				& $^{2}$H  & 2.42E-09 & 1.36E-16\\
				& $^{3}$He & 5.20E-09 & 3.20E-04\\
				& $^{4}$He & 1.13E-02 & 3.39E+00\\
				& $^{6}$Li & 1.43E-13 & 4.12E-13\\
				& $^{7}$Li & 4.41E-11 & 2.92E-10\\
				& $^{9}$Be & 1.24E-11 & 2.74E-11\\
				& $^{10}$B & 1.63E-11 & 8.00E-10\\
				& $^{11}$B & 1.10E-08 & 2.48E-09\\
				& $^{12}$C & 5.16E-04 & 9.27E-02\\
			\tableline
		\end{tabular}
	\end{center}
    \tablecomments{Table \ref{tab:finabw07} is published in its entirety in the machine-readable format. A portion is shown here for guidance regarding its form and content.}
	%\tablecomments{The table columns are: zero age main sequence \textbf{(ZAMS)} mass, compactness at bounce, total mass at collapse, mass of the iron core, carbon-oxygen core, and helium core, mass of the hydrogen-rich envelope, explosion energy ($1\; {\rm B}= 1\;{\rm Bethe} = 10^{51}~\rm erg$) and remnant mass. The column after the remnant mass indicates which layer in the pre-explosion structure this corresponds to. The last four columns give the nucleosynthesis yields of observable isotopes of nickel and titanium. 
%	}
\end{table*}

\end{document}